\documentclass[acmsmall,nonacm,natbib=false,11pt]{acmart}
\usepackage{float}
\usepackage{algorithm}
\usepackage{algpseudocode}
\usepackage{supertabular}
\usepackage[backend=biber,datamodel=acmdatamodel,style=acmnumeric,sortcites=true]{biblatex}
\DeclareSourcemap{
  \maps[datatype=bibtex]{
    \map{
      \step[fieldsource=doi,
        match=\regexp{(?i)\A(?:https?://(?:dx\.)?doi\.org/|doi:\s*)},
        replace={}]
    }
    \map{
      \step[fieldsource=title,
        matchi=\regexp{\bCOMB\b(?!\})},
        replace=\regexp{\{COMB\}}]
      \step[fieldsource=title,
        matchi=\regexp{\bPET\b(?!\})},
        replace=\regexp{\{PET\}}]
      \step[fieldsource=title,
        matchi=\regexp{\bGaussian\b(?!\})},
        replace=\regexp{\{Gaussian\}}]
      \step[fieldsource=title,
        matchi=\regexp{\bBayesian\b(?!\})},
        replace=\regexp{\{Bayesian\}}]
      \step[fieldsource=title,
        matchi=\regexp{\bMonte\s+Carlo\b(?!\})},
        replace=\regexp{\{Monte\x20Carlo\}}]
      \step[fieldsource=title,
        matchi=\regexp{\bFrank-Wolfe\b(?!\})},
        replace=\regexp{\{Frank-Wolfe\}}]
      \step[fieldsource=title,
        matchi=\regexp{\bDeal\s+or\s+No\s+Deal\b(?!\})},
        replace=\regexp{\{Deal\x20or\x20No\x20Deal\}}]
    }
    \map{
      \step[fieldsource=entrykey,
        match=\regexp{\Akuhn1953extensive\z}, final]
      \step[typesource=inbook, typetarget=incollection]
      \step[fieldsource=title,
        match=\regexp{\A11\.\s+Extensive},
        replace=\regexp{11.\x20\{Extensive\}}]
    }
    \map{
      \step[fieldsource=entrykey,
        match=\regexp{\A(?:southey2005bayes|ganzfried2011opponent|williams2011gaussian|ponsen2008bayes|albrecht2013hba)\z},
        final]
      \step[fieldsource=location, fieldtarget=venue]
      \step[fieldsource=address, fieldtarget=location]
    }
    \map{
      \pertype{inproceedings}
      \step[fieldsource=publisher,
        match=\regexp{\A(?:PMLR|Association\x20for\x20Computational\x20Linguistics)\z},
        final]
      \step[fieldsource=address, fieldtarget=venue]
    }
    \map{
      \step[fieldsource=date,
        match=\regexp{\A(\d{4})/(\d{2})/(\d{2})\z},
        replace=\regexp{$1-$2-$3}]
    }
    \map{
      \pertype{article}
      \step[fieldsource=isbn,
        match=\regexp{\A\d{4}-\d{3}[\dXx]\z}, final]
      \step[fieldsource=isbn, fieldtarget=issn]
    }
    \map{
      \step[fieldsource=month,
        match=\regexp{\A\d+--\d+\s+Apr\z}, replace={4}]
      \step[fieldsource=month,
        match=\regexp{\A\d+--\d+\s+Jul\z}, replace={7}]
    }
  }
}

\renewbibmacro*{date}{%
  \ifboolexpr{
    test {\ifentrytype{misc}} and
    test {\iffieldundef{month}} and
    (test {\iffieldequalstr{eprinttype}{arXiv}} or
     test {\iffieldequalstr{eprinttype}{arxiv}})
  }{}{\printtext[parens]{\printdate}}%
}

\AtEveryBibitem{%
  \clearfield{issn}%
  \clearfield{eprintclass}%
  \ifboolexpr{
    test {\ifentrytype{article}} or
    test {\ifentrytype{inproceedings}} or
    test {\ifentrytype{incollection}} or
    test {\ifentrytype{inbook}}
  }{%
    \clearfield{month}%
    \clearfield{day}%
    \iffieldundef{doi}{}{\clearfield{eprint}}%
  }{}%
  \ifentrytype{inproceedings}{%
    \clearname{editor}%
    \iffieldundef{volume}{\clearfield{series}}{}%
    \clearfield{venue}%
    \ifboolexpr{
      test {\iffieldequalstr{entrykey}{brown2019deepcfr}} or
      test {\iffieldequalstr{entrykey}{ge2024safe}} or
      test {\iffieldequalstr{entrykey}{jaggi2013revisiting}} or
      test {\iffieldequalstr{entrykey}{johanson2009data}}
    }{\clearfield{url}}{}%
  }{}%
  \iflistsequal{organization}{publisher}{\clearlist{organization}}{}%
  \ifboolexpr{
    test {\iffieldequalstr{note}{Main Track}} or
    test {\iffieldequalstr{note}{Multi-Agent Rationality}} or
    test {\iffieldequalstr{note}{International Conference on MultiAgent Systems 2000}}
  }{\clearfield{note}}{}%
  \ifboolexpr{
    test {\iffieldequalstr{eprinttype}{arXiv}} or
    test {\iffieldequalstr{eprinttype}{arxiv}}
  }{%
    \iffieldundef{eprint}{}{%
      \clearfield{url}%
      \iffieldequalstr{journaltitle}{CoRR}{\clearfield{volume}}{}%
    }%
  }{}%
}
\setcopyright{none}
\startPage{0}
\fancypagestyle{preprint}{%
  \fancyhf{}%
  \fancyhead[C]{\footnotesize\shortauthors}%
  \fancyhead[R]{\thepage}%
}
\acmConference[WWW '27]{The ACM Web Conference 2027}{May 10--14, 2027}{Dublin, Ireland}
\acmYear{2027}
\copyrightyear{2027}
\acmDOI{}
\acmISBN{}

\newcommand{\KL}{D_{\mathrm{KL}}}

\newcommand{\p}[1]{\left( #1\right)}
\newcommand{\brac}[1]{\left\{ #1\right\}}

\newtheorem{theorem}{Theorem}
\newtheorem{lemma}{Lemma}
\newtheorem{appendixlemma}{Lemma}[section]
\newtheorem{corollary}{Corollary}

\ifPDFTeX
  \DeclareFontFamily{T1}{ourmethod}{}
  \DeclareFontShape{T1}{ourmethod}{m}{n}{<-> s*[1.1687] ec-qcrr}{}
  \newcommand{\ourmethodfont}{\fontencoding{T1}\fontfamily{ourmethod}\selectfont}
\else
  \newfontfamily\ourmethodfont[Scale=MatchUppercase]{texgyrecursor-regular.otf}
\fi

\DeclareRobustCommand{\SLME}{\ourmethod{SLME}}
\DeclareRobustCommand{\OISR}{\ourmethod{OISR}}
\DeclareRobustCommand{\BBR}{\baselinemethod{BBR}}

\DeclareRobustCommand{\MSEM}{\baselinemethod{MS-EM}}
\DeclareRobustCommand{\FMAP}{\baselinemethod{FMAP}}

\DeclareRobustCommand{\chance}{\ifmmode\mathrm{chance}\else\textup{chance}\fi}

\title{Scalable Opponent Strategy Estimation in Imperfect-Information Games}
\makeatletter
\patchcmd{\@mktitle@i}{\raggedright}{\centering}{}{}
\patchcmd{\@starttoc}{\addvspace{32\p@\@plus14\p@}}
  {\addvspace{\medskipamount}}{}{}
\makeatother
\author{Tri-An D. Nguyen}
\affiliation{%
  \institution{Rutgers University}, New Brunswick
  \city{New Brunswick}
  \country{USA}
}
\author{Garrett Seo}
\affiliation{%
  \institution{University of Michigan}
  \city{Ann Arbor}
  \country{USA}
}
\author{Xintong Wang}
\affiliation{%
  \institution{Rutgers University}, New Brunswick
  \city{New Brunswick}
  \country{USA}
}
\authorsaddresses{}
\begin{document}
\begin{abstract}

Repeated interactions on the Web provide agents with a growing record of other agents' actions from which to predict future behavior. However, agents observe only part of each encounter, complicating this learning task. We introduce two algorithms that estimate a fixed opponent strategy from these partial observations. Strategy Library Mixture Estimation (\SLME{}) models the opponent's strategy as a mixture of strategies from a predefined library and converges to a mixture that best explains the observations. When the library covers the styles of play an opponent might use, such as different negotiation tactics, \SLME{} models opponents who play a mixed strategy over these styles. Observation-Induced Strategy Reconstruction (\OISR{}) solves a maximum-a-posteriori problem using only the histories compatible with the observations rather than the full game tree. We define \emph{observational equivalence} to describe opponent strategies that observations cannot distinguish. Such strategies have the same best responses, which give the learner the same expected payoff. As observations accumulate under sufficient exploration, \OISR{}'s model becomes \emph{asymptotically equivalent} to the opponent's strategy, and \SLME{}'s model does so when the opponent is representable, meaning its strategy is equivalent to a mixture of library strategies. We evaluate \SLME{} and \OISR{} against three baselines in Leduc poker and Deal or No Deal. Given as many observed plays as the opponent has information sets, \SLME{} reaches normalized regret below 0.01 against representable opponents, lower than the evaluated library-free methods reach with $100\times$ as many plays. \OISR{} reaches regret comparable to that of the only tested baseline that also guarantees asymptotic equivalence for any fixed opponent, while fitting $269\times$ faster on average when the learner acts first and $32\times$ faster when it acts second.

\end{abstract}


\renewcommand{\keywordsname}{Keywords}
\raggedbottom
\maketitle
\thispagestyle{empty}
\setcounter{tocdepth}{1}
\tableofcontents
{\small\noindent\textbf{AI Disclosure}\par\nobreak
\noindent We used AI coding assistants (Claude Code and OpenAI Codex) to help explore related work, implement code, and check the proofs.\par}
\clearpage
\pagestyle{preprint}
\section{Introduction}
\label{sec:introduction}

The growing use of automated negotiators, bidding agents, and other algorithms acting on behalf of users enables repeated economic interactions on the Web at scale. Opponent modeling uses these interactions to predict how other agents will act~\cite{albrecht2018survey}, enabling an agent, which we call the learner, to best-respond. In negotiation, for example, the other party's previous offers and rejections can inform which future offers to make and which to accept.

We study how to model an opponent's strategy from repeated play in finite, two-player, general-sum games with imperfect information, assuming a fixed opponent strategy and perfect recall. This setting poses two key challenges. First, the learner may observe only part of each play, so the context behind the opponent's actions can be hidden: an offer may be rejected because of the opponent's private valuation or because of its negotiation strategy. Second, the opponent's strategy specifies how to act at each of its information sets, whose number typically grows exponentially with the number of turns. Existing methods face complementary limitations. Library-based methods hypothesize that the opponent's strategy is one of the members of a finite strategy library~\cite{southey2005bayes,ganzfried2018bayesian,ganzfried2026consistent} and thus can fail when no single member closely matches the opponent. Full-game maximum-a-posteriori estimation (\FMAP{})~\cite{ganzfried2026consistent} guarantees asymptotic observational equivalence to the opponent's strategy, but its cost grows with the number of opponent information sets.

We address these limitations with two methods. Strategy Library Mixture Estimation (\SLME{}) treats library strategies as building blocks rather than competing hypotheses. It models the opponent's strategy as a mixture of library strategies and finds a mixture that maximizes the observed-data likelihood, and thus does not require any single library member to closely match the opponent. When the library covers the styles of play an opponent might use, such as established negotiation tactics, \SLME{} models opponents who play a mixed strategy over these styles. Observation-Induced Strategy Reconstruction (\OISR{}) estimates the opponent's strategy without a library. It solves a compact reformulation of the full-game maximum-a-posteriori (MAP) problem, built from only the hidden histories compatible with past observations: it retains the opponent's information sets along these histories and sets the action distributions at the remaining information sets to the modes of a fixed prior.

We prove that, with the initializations and priors we construct, \SLME{} and \OISR{} converge to the global optima of their objectives with explicit bounds on the remaining optimization gap. We define \emph{observational equivalence} to describe opponent strategies that observations cannot distinguish. Such strategies have the same best responses, which give the learner the same expected payoff. As observations accumulate, provided the learner explores sufficiently and the per-play optimization gap vanishes, \OISR{}'s model becomes \emph{asymptotically equivalent} to the opponent's strategy, and \SLME{}'s model does so when the opponent is representable, meaning its strategy is equivalent to a mixture of library strategies. Both guarantees hold even when the learner updates its strategy between plays.

We evaluate \SLME{} and \OISR{} against three baselines in Leduc poker and two sizes of Deal or No Deal, small and large. We test them offline, where all methods fit the same observed plays, and online in Leduc, where the learner refits its model after each play. We report normalized regret, which divides the learner's regret by that of a best response to an uninformative uniform opponent model, so that results are comparable across games with different payoff scales. Offline, \SLME{} reaches normalized regret below 0.01 against representable opponents once the number of observed plays equals the opponent's number of information sets, lower than library-free methods reach with $100\times$ as many plays. Against unrepresentable opponents, the library-based methods do no better than the uniform model, while \OISR{} improves at larger dataset sizes, reaching normalized regret between 0.021 and 0.058 with 100 observed plays per opponent information set. In Leduc and small Deal or No Deal, where we also evaluate \FMAP{}, \OISR{} attains regret comparable to \FMAP{}'s while fitting $269\times$ faster on average when the learner acts first and $32\times$ faster when it acts second. Online, \SLME{} falls below 0.01 normalized regret within 110 plays against representable opponents, and against unrepresentable opponents \OISR{} averages 0.49, comparable to 0.54 for \FMAP{} and well below 1.6--3.9 for the other methods.
\subsection{Related Work}
\label{sec:related-work}

\paragraph{Bayesian and library-based estimation.}
Bayesian approaches infer a posterior over opponent strategies and compute responses to it~\cite{southey2005bayes,ganzfried2018bayesian}.
Related methods learn opponent-specific predictions from poker observations with unrevealed cards~\cite{ponsen2008bayes}, or select actions using posterior beliefs over candidate behavioral types~\cite{albrecht2013hba}.
Finite-library implementations of Bayesian best response (BBR) treat each library strategy as a hypothesis about the opponent's fixed strategy across all observed plays~\cite{southey2005bayes}.
Their posterior weights can concentrate on a single library strategy even when a mixture of library strategies represents the opponent exactly~\cite{ganzfried2026consistent}.
\SLME{} treats the library weights as mixture-model parameters and estimates them by maximizing the observed-data likelihood, directly targeting the best-fitting library mixture while preserving the library strategies' behavioral structure.

\paragraph{Direct strategy estimation.}
Southey et al.\ derive MAP estimates of opponent strategy parameters in Kuhn poker~\cite{southey2009effective}.
Mealing and Shapiro combine expectation-maximization (EM) inference of hidden information with sequence prediction and simulated play to model changing opponents in simplified poker~\cite{mealing2017em}.
Ganzfried's \FMAP{} maximizes a concave log posterior over the full opponent sequence-form representation using projected gradient
steps~\cite{ganzfried2026consistent}.
Under two assumptions---that no other strategy is observationally equivalent to the opponent's and that play visits every opponent information set infinitely often---\FMAP{}'s model converges to the opponent's strategy~\cite{ganzfried2026consistent}.
Trading these assumptions for a sufficiently explorative learner, our analysis (Section~\ref{sec:statistical-guarantees}) shows that \FMAP{}'s model becomes asymptotically observationally equivalent to the opponent's strategy.
However, \FMAP{} optimizes over all of the opponent's information sets, even those that no observed play could have visited.
\OISR{} instead solves a full-game MAP problem while representing only the opponent's information sets along compatible histories.

\paragraph{Equilibrium computation and opponent exploitation.}
Methods for approximating Nash equilibria in two-player zero-sum games include counterfactual regret minimization and its sampled and neural variants~\cite{zinkevich2007cfr,lanctot2009mccfr,brown2019deepcfr}, as well as neural fictitious self-play~\cite{heinrich2016nfsp}.
Opponent exploitation adapts play using explicit opponent models~\cite{ganzfried2011opponent} or estimated utilities of precomputed responses~\cite{bard2013implicit}.
Robust response methods balance exploitation against exploitability~\cite{johanson2007robust,johanson2009data}, while Opponent Exploitation Search (OX-Search) controls exploitability relative to a baseline~\cite{ge2024safe}.
Our work addresses the complementary task of estimating opponent strategies from partial observations for subsequent response computation.

\paragraph{Automated negotiation.}
Opponent models support offer generation, concession, and acceptance decisions~\cite{jennings2001negotiation,baarslag2016opponent}.
Bayesian approaches infer private preferences~\cite{hindriks2008bayesian,chang2024comb}, while Gaussian process models predict the utility of future opponent offers~\cite{williams2011gaussian}.
We estimate behavioral strategies without requiring knowledge of opponent utilities and evaluate our methods in two sizes of the Deal or No Deal negotiation game~\cite{lewis2017deal}.

\section{Problem Formulation}
\label{sec:problem-formulation}

\paragraph{Game formulation}
We consider finite, two-player, general-sum extensive-form games with imperfect information and perfect recall. 
Let $i$ denote the learner and $-i$ the opponent. 
Private information, such as a player's hand in poker or valuation in negotiation, is represented by chance moves whose outcomes are privately observed.

A history $h$ is a sequence of player actions and chance outcomes from the start of a play.
Let $H$ denote the set of histories and $Z \subseteq H$ the set of terminal histories. 
At each nonterminal history $h$, $P(h)\in \{i,-i,\chance{}\}$ identifies who acts next, and $A(h)$ is the set of legal actions. %
At a terminal history $z$, the learner receives utility $u_i(z)\in\mathbb{R}$.
The learner knows the game rules (e.g., information structure and chance probabilities) and its own utility function; knowledge of the opponent's utility is not required.
We write $p_{\chance{}}(h)$ for the product of the probabilities of all chance outcomes along history $h$, with value one if there are no chance moves.

Let $o_i$ be a function mapping each history $h$ to the information available to player $i$. 
Under perfect recall, $o_i(h)$ contains the learner's complete record of observations and its own actions along $h$. We assume that the learner observes its utility at the end of each play: at a terminal history $z$, $o_i(z)$ includes $u_i(z)$.
Let $\mathcal{I}_i$ denote the collection of player $i$'s information sets, which partitions the histories at which player $i$ acts.
Two histories $h$ and $h'$ at which the learner acts belong to the same information set $I$ exactly when $o_i(h)=o_i(h')$.
All histories in $I$ have the same set of legal actions, denoted $A(I)$.

\paragraph{Strategies and realization weights}
A \textit{behavioral strategy} $\sigma_i$ specifies an action distribution at each information set $I\in \mathcal{I}_i$~\cite{kuhn1953extensive}.
We write $\sigma_i(a\mid I)$ for the probability of choosing action $a \in A(I)$.
Let $\Sigma_i$ denote the set of behavioral strategies for the learner.
For each history $h$, let $s_i(h)$ record player $i$'s decisions along history $h$ as an ordered sequence of pairs $(I, a)$, where $I$ is the information set and $a$ is the action chosen there.
Let $S_i$ be the set of all such sequences, including the empty sequence $\emptyset$, used when player $i$ has not yet acted. Under perfect recall, all histories in an information set $I$ share the same sequence of the player's previous information sets and chosen actions. We call this the \textit{preceding sequence} of $I$ and denote it by $\rho(I)$.
For any sequence $s\in S_i$, define its \textit{realization weight} under $\sigma_i$ as
\[y_{i,s}(\sigma_i):=\prod_{(I,a)\in s}\sigma_i(a\mid I), \qquad y_{i,\emptyset}(\sigma_i)=1,\]
i.e., the product of player $i$'s own action probabilities along $s$.
We use $y_i(\sigma_i)=(y_{i,s}(\sigma_i))_{s\in S_i}$ to denote the vector of these weights.
For $I\in\mathcal{I}_i$ and $a\in A(I)$, write $Ia$ for the sequence $\rho(I)$ followed by $(I,a)$. Since the action probabilities at $I$ sum to one, realization weights satisfy the \textit{flow constraints}
\[
\sum_{a\in A(I)}y_{i,Ia}(\sigma_i)=y_{i,\rho(I)}(\sigma_i)\qquad\text{for every }I\in\mathcal{I}_i.
\]
Conversely, every nonnegative vector $y$ with $y_{\emptyset}=1$ that satisfies these constraints is the weight vector of any strategy that plays $a$ at $I$ with probability $y_{Ia}/y_{\rho(I)}$ wherever $y_{\rho(I)}>0$. The set $Y_i$ of realization-weight vectors is therefore a polytope; this representation is known as the \textit{sequence form}~\cite{vonstengel1996efficient}.
We define $\Sigma_{-i}$, $s_{-i}(h)$, $\mathcal{I}_{-i}$, $S_{-i}$, $y_{-i}(\sigma_{-i})$, and $Y_{-i}$ analogously for the opponent.

\paragraph{Observations and estimation objective}
Our goal is to estimate the unknown opponent strategy, represented as a behavioral strategy $\sigma^*_{-i}$,\footnote{By Kuhn's theorem~\cite{kuhn1953extensive}, under perfect recall, every mixed strategy has a behavioral strategy that induces the same distribution over terminal histories against every strategy of the other player. Thus, representing the opponent's strategy in behavioral form entails no loss of generality in terms of game outcomes.} from $n$ observed plays of a fixed game.
Following prior work on opponent modeling~\cite{southey2005bayes,ganzfried2018bayesian,ganzfried2026consistent}, we assume that the opponent's strategy remains fixed across plays.
In play $t$, let $z^t \in Z$ be the resulting terminal history, and $\sigma_i^t$ the learner's behavioral strategy, which may depend on observations from earlier plays. 
The learner observes $o_i(z^t)$, which may be compatible with several terminal histories:
\[
Z^t := \{z \in Z : o_i(z) = o_i(z^t)\}.
\]
As the observation $o_i(z)$ includes the learner's past observations and actions, all histories $z \in Z^t$ have the same learner sequence $s_i(z)=s_i(z^t)$.
For a candidate opponent strategy $\sigma_{-i}$, the probability of observing $o_i(z^t)$, conditional on the learner strategy in play $t$, is therefore
\[
\begin{aligned}
q^t(\sigma_{-i})&:=\sum_{z\in Z^t}y_{i,s_i(z)}(\sigma_i^t)\,p_{\mathrm{chance}}(z)\,y_{-i,s_{-i}(z)}(\sigma_{-i})\\
&\phantom{:}=y_{i,s_i(z^t)}(\sigma_i^t)\,f_{o_i(z^t)}(\sigma_{-i}),
\end{aligned}
\]
where, for each $o\in U:=o_i(Z)$, the \emph{opponent factor} $f_o(\sigma_{-i}):=\sum_{z\in Z:\,o_i(z)=o}p_{\mathrm{chance}}(z)\,y_{-i,s_{-i}(z)}(\sigma_{-i})$ is the probability of $o$ with the learner's realization weight divided out. We call two opponent strategies \textit{observationally equivalent} if they have the same $f_o$ for every $o\in U$. By the factorization, observationally equivalent strategies induce the same distribution of learner observations under every learner strategy.

Let $u_i(\sigma_i,\sigma_{-i})$ denote the learner's expected utility when it plays $\sigma_i$ and the opponent plays $\sigma_{-i}$, and $\mathrm{BR}(\sigma_{-i})$ a best response to $\sigma_{-i}$, a learner strategy that maximizes $u_i(\cdot,\sigma_{-i})$, with ties broken by a fixed rule. The \textit{regret} of a strategy $\sigma_i$ is the expected utility the learner loses against the opponent by playing $\sigma_i$ instead of best-responding to the opponent's strategy. We define the regret of a model $\widehat{\sigma}_{-i}$ as the regret of the best response to $\widehat{\sigma}_{-i}$:
\begin{equation}
R(\widehat{\sigma}_{-i}):=u_i\bigl(\mathrm{BR}(\sigma_{-i}^*),\sigma_{-i}^*\bigr)-u_i\bigl(\mathrm{BR}(\widehat{\sigma}_{-i}),\sigma_{-i}^*\bigr).
\label{eq:regret}
\end{equation}
Since the learner's observation includes its utility, each learner strategy earns the same expected utility against any two observationally equivalent strategies, so every model observationally equivalent to the opponent's strategy has zero regret. Thus, as the number of plays grows, we seek fitted models whose values of $f_o$ approach those of the opponent's strategy for every $o\in U$.

Both methods use the likelihood of the recorded observations. To compute these likelihoods efficiently, we group plays whose likelihoods are identical. For each play $t$ and compatible terminal history $z\in Z^t$, define the \textit{known coefficient}
\[
c_z^t
:=
y_{i,s_i(z)}(\sigma_i^t)\,p_{\mathrm{chance}}(z),
\]
which does not depend on the candidate opponent strategy, so that $q^t(\sigma_{-i})=\sum_{z\in Z^t}c_z^t\,y_{-i,s_{-i}(z)}(\sigma_{-i})$. We group plays into the same \textit{observation type} when they yield the same learner observation and have the same collection of known coefficients. These plays have identical likelihoods under every candidate opponent strategy.
Let $V$ denote the recorded observation types and $n_j$ the number of plays of type $v_j\in V$. Plays of type $v_j$ have the same compatible histories $Z_j$ and known coefficients $c_{j,z}$, so their common likelihood under candidate $\sigma_{-i}$ is $q_j(\sigma_{-i})=\sum_{z\in Z_j}c_{j,z}\,y_{-i,s_{-i}(z)}(\sigma_{-i})$.
The observed-data log likelihood is
\begin{equation}
\ell_n(\sigma_{-i})
:=
\sum_{t=1}^{n}\log q^t(\sigma_{-i})
=
\sum_{j=1}^{|V|}n_j\log q_j(\sigma_{-i}).
\label{eq:data-log-likelihood}
\end{equation}

\section[Strategy Library Mixture Estimation (SLME)]{Strategy Library Mixture Estimation}
\label{sec:slme}
We first consider estimation based on a fixed library of candidate opponent strategies.
For example, such a library can incorporate domain knowledge through representative behavioral patterns, encoding concession patterns in negotiation~\cite{faratin1998negotiation} or bluffing frequencies in poker~\cite{southey2005bayes}.
\SLME{}  directly estimates mixture weights by maximizing the observed-data likelihood, allowing multiple library members to jointly explain the observations.

\paragraph{Mixture model}
Let $\{\sigma_{-i}^m\}_{m=1}^{M}$ be the library of opponent behavioral strategies.
We model the opponent as independently selecting a library member at the beginning of each play and following it throughout that play.
The selection probabilities $\pi\in\Delta_M$ remain fixed across plays, where $\Delta_M$ is the probability simplex over the $M$ members and $\pi_m$ is the probability of selecting library member $m$.

For observation type $v_j$, let $q_j^m:=q_j(\sigma_{-i}^m)$ denote its likelihood under member $m$, as defined in Section~\ref{sec:problem-formulation}.
Its likelihood under the mixture is
\[
q_j(\pi)=\sum_{m=1}^M \pi_m q_j^m.
\]
We estimate the mixture weights by maximizing the observed-data log likelihood:
\begin{equation}
\widehat{\pi}
\in
\operatorname*{arg\,max}_{\pi\in\Delta_M}
\ell_n(\pi),
\qquad
\ell_n(\pi)
=
\sum_{j=1}^{|V|}n_j\log q_j(\pi).
\label{eq:slme-mle}
\end{equation}
The objective is concave in $\pi$ because each $q_j(\pi)$ is linear in $\pi$.

\paragraph{Fitting and convergence}
We fit the mixture weights using the EM algorithm, treating the selected library member in each play as a latent variable~\cite{dempster1977em}.
At iteration $k$, the E-step computes the responsibility
\[
\gamma_j^{m,(k)}
=
\frac{\pi_m^{(k)}q_j^m}
{\sum_{r=1}^M \pi_r^{(k)}q_j^r},
\]
the conditional probability that member $m$ generated an observation of type $v_j$.
The M-step averages these responsibilities over the recorded plays:
\[
\pi_m^{(k+1)}
=
\frac{1}{n}\sum_{j=1}^{|V|}n_j\gamma_j^{m,(k)}.
\]
Each iteration uses the entire accumulated dataset.
We have the following guarantee for exact EM, whose iterations are computed without numerical error, as a direct application of Zhao's convergence analysis~\cite{zhao2022multiplicative}.

\begin{lemma}[{\SLME{} convergence}]
\label{lem:slme-rate}
Suppose $n\geq 1$ and every observation type has positive likelihood under at least one library member. 
Starting from $\pi_m^{(0)}=1/M$ for all $m$, the exact EM iterates converge to a maximum-likelihood solution $\widehat{\pi}$ and satisfy
\begin{equation}
0\leq\frac{\ell_n(\widehat{\pi})-\ell_n(\pi^{(k)})}{n}\leq\frac{\log M}{k+1}.
\label{eq:slme-em-rate}
\end{equation}
\end{lemma}
\begin{proof}
Let $f(\pi):=\ell_n(\pi)/n$.
The EM update satisfies $\pi_m^{(k+1)}=\pi_m^{(k)}\nabla_m f(\pi^{(k)})$, so it coincides with the multiplicative gradient method.
The objective has Zhao's required form: the coefficients $n_j/n$ are positive and sum to one, and each likelihood vector $(q_j^1,\ldots,q_j^M)$ is nonnegative and nonzero.
Zhao's convergence and rate guarantees therefore apply.
\end{proof}
Lemma~\ref{lem:slme-rate} guarantees convergence to the maximum log likelihood over library mixtures for observations from a fixed set of plays.

\paragraph*{Computation}
We cache the log likelihood $\log q_j^m$ of each observation type $v_j$ under each library strategy $\sigma_{-i}^m$. 
For each observation type, we enumerate its compatible terminal histories once and sum their probabilities under each library strategy. %
The likelihood table requires $O(|V|M)$ numerical entries, with $O(|V|)$ additional entries for the counts $n_j$.
After preparation, each EM iteration on a fixed dataset of plays costs $O(|V|M)$ scalar operations and requires no further history enumeration or evaluation of library strategies.

For online use, a new play either increments the count of a recorded type or adds a new type, whose likelihood under each library strategy is computed once; each refit then runs EM on all accumulated types at $O(|V|M)$ per iteration, whereas finite-library BBR updates its posterior in $O(M)$ operations per play~\cite{southey2005bayes}. Appendix~\ref{app:slme-computation} analyzes the time and space costs of \SLME{}.
\section[Observation-Induced Strategy Reconstruction (OISR)]{Observation-Induced Strategy Reconstruction}
\label{sec:oisr}
When a strategy library is unavailable or cannot adequately represent the opponent's behavior, \OISR{} estimates the opponent's behavioral strategy directly, using a compact representation constructed from the recorded observations.
It optimizes action distributions at opponent information sets appearing in compatible terminal histories and fixes the remaining distributions at their prior modes.
This reduction preserves the full MAP optimum for the chosen prior while avoiding explicit representation of the information sets outside the compatible histories, whose action distributions do not affect the observed-data likelihood.

\paragraph{Objective and prior}
We place fixed, independent Dirichlet priors on the opponent's action distributions at its information sets. 
For each $I \in \mathcal{I}_{-i}$ and $a\in A(I)$, let $\lambda(I,a)\geq0$ be the pseudocount, so that the Dirichlet shape parameter is $\alpha(I,a)=1+\lambda(I,a)$. 
We call the prior \textit{positive} if every pseudocount is positive. 
Adding the log-prior terms to the observed-data log likelihood in Eq.~\eqref{eq:data-log-likelihood} gives the MAP objective, up to an additive constant independent of the strategy:
\begin{equation}
J_n(\sigma_{-i}) := \ell_n(\sigma_{-i}) + \sum_{\substack{I\in\mathcal{I}_{-i}\\a\in A(I)}}\lambda(I,a)\log \sigma_{-i}(a\mid I).
\label{eq:oisr-objective}
\end{equation}

\paragraph{Compact representation}
Let
$\widetilde{\mathcal{I}}_{-i}$ be the set of all opponent information sets appearing in at least one compatible terminal history $z\in Z_j$ for some recorded type $v_j\in V$.
We call these information sets \textit{retained} and assume that $\widetilde{\mathcal{I}}_{-i}\neq\emptyset$; otherwise the observed plays contain no opponent decisions to fit.

Assume a positive prior, and write
$\lambda(I):=\sum_{a\in A(I)}\lambda(I,a)$.
At a non-retained information set, the action distribution does not affect the observed-data likelihood and appears only in its own log-prior term, maximized at the prior mode, $\sigma_{-i}(a\mid I)=\lambda(I,a)/\lambda(I)$.
\OISR{} fixes these distributions at their prior modes, making their total log-prior contribution a constant $K$ and leaving only the retained action distributions to optimize:
\begin{equation}
J_n(\sigma_{-i})
=
\ell_n(\sigma_{-i})
+
\sum_{\substack{I\in\widetilde{\mathcal{I}}_{-i}\\a\in A(I)}}
\lambda(I,a)\log\sigma_{-i}(a\mid I) +K,
\label{eq:oisr-compact-objective}
\end{equation}
where $K:=\sum_{I\in\mathcal{I}_{-i}\setminus\widetilde{\mathcal{I}}_{-i}}
\sum_{a\in A(I)}
\lambda(I,a)\log\bigl(\lambda(I,a)/\lambda(I)\bigr)$. 
Optimizing the retained distributions and completing the rest at their prior modes therefore solves the full MAP problem exactly.

Let $\widetilde{S}_{-i}:=\{\emptyset\}\cup
\{Ia:I\in\widetilde{\mathcal{I}}_{-i},\,a\in A(I)\}$ be the retained sequences, and let $\widetilde{Y}_{-i}$ be the set of restrictions of realization-weight vectors in $Y_{-i}$ to these sequences.
For $y_{-i}\in\widetilde{Y}_{-i}$ with positive entries, the retained action probabilities are $\sigma_{-i}(a\mid I)=y_{-i,Ia}/y_{-i,\rho(I)}$.
We write $J_n(y_{-i})$ for the objective of this strategy, completed with the prior modes at non-retained information sets.

\paragraph{Fitting and monotonicity}
Grouping each type's compatible terminal histories by the opponent's sequence gives the sparse coefficient matrix $C$:
\begin{equation}
C_{j,s}:=\sum_{\substack{z\in Z_j\\s_{-i}(z)=s}}c_{j,z},
\qquad
q_j(y_{-i})=\sum_{s\in\widetilde{S}_{-i}}C_{j,s}y_{-i,s}.
\label{eq:oisr-coefficients}
\end{equation}
We call $s$ an \textit{endpoint} if $C_{j,s}>0$ for some $j$: it records the opponent's complete sequence of decisions along a compatible terminal history.

\OISR{} fits the retained action distributions by EM, treating the opponent's sequence in each recorded play as unobserved.
Let $\sigma_{-i}^{(k)}$ be the opponent model after $k$ EM iterations, and let $y_{-i}^{(k)}$ be its realization weights on $\widetilde{S}_{-i}$.
For a current model assigning positive probability to every recorded type, the E-step computes
\[
\gamma_{j,s}^{(k)}=
\frac{C_{j,s}y_{-i,s}^{(k)}}
     {q_j(y_{-i}^{(k)})},
\]
the conditional probability that a play of type $v_j$ has opponent sequence $s$.
With $n_j$ recorded plays of type $v_j$, the expected number of times the opponent chose $a$ at $I$ is
\[
N^{(k)}(I,a)
=
\sum_j n_j
\sum_{\substack{s\in\widetilde{S}_{-i}\\(I,a)\in s}}
\gamma_{j,s}^{(k)}.
\]

The M-step maximizes the expected complete-data log posterior separately at each retained information set.
With $N^{(k)}(I):=\sum_{a\in A(I)}N^{(k)}(I,a)$, the resulting update is
\begin{equation}
\sigma_{-i}^{(k+1)}(a\mid I)
=
\frac{N^{(k)}(I,a)+\lambda(I,a)}
     {N^{(k)}(I)+\lambda(I)}.
\label{eq:oisr-em-update}
\end{equation}
Thus, each update normalizes the expected action counts augmented by the prior pseudocounts.
Non-retained distributions remain fixed at their prior modes. Appendix~\ref{app:oisr-em-monotonicity} shows that
\begin{equation}
\begin{aligned}
\small
J_n(\sigma_{-i}^{(k+1)})-J_n(\sigma_{-i}^{(k)})
&\geq
\sum_{\substack{I\in\widetilde{\mathcal{I}}_{-i}\\a\in A(I)}}
\bigl(N^{(k)}(I,a)+\lambda(I,a)\bigr)
\log
\frac{\sigma_{-i}^{(k+1)}(a\mid I)}
     {\sigma_{-i}^{(k)}(a\mid I)}\\
&\geq 0,
\end{aligned}
\label{eq:oisr-em-bound}
\end{equation}
so no EM iteration decreases the objective.

\paragraph{Optimization guarantees}
The update in Eq.~\eqref{eq:oisr-em-update} and the bound in Eq.~\eqref{eq:oisr-em-bound} hold for any positive pseudocounts.
For \textit{global convergence}, we impose a sufficient condition ensuring that the log-prior contribution is concave in realization weights.
We call the prior \emph{coherent} if
\begin{equation}
\lambda(I,a)
\geq
\sum_{\substack{I'\in\mathcal{I}_{-i}\\\rho(I')=Ia}}
\lambda(I')
\qquad
\text{for every } I\in\mathcal{I}_{-i},\ a\in A(I).
\label{eq:oisr-coherence}
\end{equation}
Here $I'$ ranges over all possible next opponent information
sets after action $a$ at $I$.
Intuitively, coherence constrains the relative strengths of the local priors: 
the opponent reaches one of these next information sets only after choosing $a$ at $I$, so coherence requires their total pseudocount to be at most $\lambda(I,a)$.
Appendix~\ref{app:oisr-prior} constructs a prior with positive, coherent pseudocounts.

Coherence ensures $J_n$ is concave on the positive part of $\widetilde{Y}_{-i}$ (Appendix~\ref{app:oisr-retained-prior}), so every local maximum is a global maximum.
Appendix~\ref{app:oisr-em-convergence} proves the following convergence guarantee for exact EM:

\begin{theorem}[MAP optimization]
\label{thm:oisr-optimization}
Suppose $n\geq 1$, the prior is coherent with positive pseudocounts, and every row of $C$ has a positive entry.
Then exact EM, initialized with positive action probabilities at every retained information set and the prior modes elsewhere, satisfies
\[
\begin{aligned}
\lim_{k\to\infty}J_n(\sigma_{-i}^{(k)})
=J_n^*, 
\quad
\text{where } J_n^*:=\max_{\sigma_{-i}\in\Sigma_{-i}}J_n(\sigma_{-i}).
\end{aligned}
\]
\end{theorem}
This theorem guarantees convergence to the optimal MAP objective value for observations from a fixed set of plays.
We now construct \OISR{}'s initialization, which satisfies the initialization requirement of Theorem~\ref{thm:oisr-optimization} and yields an explicit convergence rate. Let
\[
x\in
\operatorname*{arg\,max}_{y_{-i}\in\widetilde{Y}_{-i}}
\min_{\text{endpoint } s}y_{-i,s},
\qquad
\Gamma:=\frac{1}{\min_{\text{endpoint } s}x_s}.
\]
Let $y_{-i}^{\mathrm{def}}$ be the realization weights, restricted to $\widetilde{S}_{-i}$, of the default strategy, which plays the prior mode at every information set, and let
$\tau:=\sum_{I\in\widetilde{\mathcal{I}}_{-i}:\rho(I)=\emptyset}
\lambda(I)$ be the total pseudocount at the opponent's first
retained decisions.
\OISR{} initializes the realization weights on $\widetilde{S}_{-i}$ to
\begin{equation}
y_{-i}^{(0)}
:=
\frac{nx+\tau y_{-i}^{\mathrm{def}}}{n+\tau}.
\label{eq:oisr-initialization}
\end{equation}
Since $\widetilde{\mathcal{I}}_{-i}\neq\emptyset$ and the prior is positive, $\tau>0$, so this combination lies in $\widetilde{Y}_{-i}$ and has positive entries, even if $x$ does not.
Appendix~\ref{app:oisr-initialization} computes $x$ and $\Gamma$ in linear time over
the retained sequences.
Appendix~\ref{app:oisr-em-rate} proves the following bound.
\begin{lemma}[Convergence rate]
\label{lem:oisr-rate}
Under the assumptions of Theorem~\ref{thm:oisr-optimization}, exact EM initialized by Eq.~\eqref{eq:oisr-initialization}
satisfies, for every $k\geq 1$,
\[
0\leq J_n^*-J_n(\sigma_{-i}^{(k)})
\leq \frac{B_0}{k},
\]
where
$B_0:=n\log\Gamma+n\log(1+\tau/n)+\tau\log(1+n/\tau)$.
\end{lemma}

\paragraph{Computation}
\label{sec:oisr-compact-representation}
For each distinct observation type, we enumerate its compatible terminal histories and aggregate their known coefficients $c_{j,z}$ by opponent sequence to construct $C_{j,s}$ (Eq.~\eqref{eq:oisr-coefficients}).
Let $E:=\brac{(j,s):C_{j,s}>0}$.
As an additional speedup, \OISR{} merges the actions that no compatible history takes at each retained information set into a single escape action carrying their total pseudocount; these actions affect $J_n$ only through their prior terms, so splitting the escape weight in proportion to their pseudocounts recovers a full MAP solution, and EM's guarantees still hold.
Using sparse likelihood calculations and backward and forward passes over the sequences $\bar S\subseteq\widetilde{S}_{-i}$ along compatible terminal histories, each EM iteration then costs $O(|V|+|E|+|\bar S|)$ scalar operations, without further history enumeration.
Appendix~\ref{app:oisr-computation} proves both claims and accounts separately for preparation costs and additional storage.

Online, a new play increments its type's count or adds a new type, whose coefficients may expand the compact representation; previously computed coefficients are reused, and each refit runs EM on all accumulated types.

\section{Statistical Guarantees}
\label{sec:statistical-guarantees}
For a fixed opponent strategy $\sigma_{-i}^*$, we call fitted models $\widehat{\sigma}_{-i,n}$ \emph{asymptotically observationally equivalent} to $\sigma_{-i}^*$ if $f_o(\widehat{\sigma}_{-i,n})\to f_o(\sigma_{-i}^*)$ for every $o\in U$ as the number of observed plays $n$ grows, the goal stated in Section~\ref{sec:problem-formulation}.
We assume sufficient exploration by the learner strategies
generating the observations: there exists $\theta>0$ such
that, almost surely, for sufficiently large $n$,
\begin{equation}
\frac{1}{n}\sum_{t=1}^{n}y_{i,s}(\sigma_i^t)
\geq\theta
\qquad\text{for every }s\in S_i.
\label{eq:statistical-coverage}
\end{equation}
A fixed learner strategy assigning positive probability
to every legal action satisfies this condition.

\paragraph{Asymptotic observational equivalence.}
The following result applies to both methods.
It requires the fitted model's log likelihood to fall below that of the opponent's strategy by at most a sublinear amount.

\begin{theorem}[Asymptotic observational equivalence]
\label{thm:opponent-recovery}
Under Eq.~\eqref{eq:statistical-coverage}, let $\widehat{\sigma}_{-i,n}$ be a model fitted from the first $n$ observed plays.
If, almost surely,
\begin{equation}
\ell_n(\widehat{\sigma}_{-i,n})
\geq \ell_n(\sigma_{-i}^*)-o(n),
\label{eq:recovery-likelihood}
\end{equation}
then, almost surely, 
$
f_o(\widehat{\sigma}_{-i,n})
\longrightarrow f_o(\sigma_{-i}^*)
\quad\text{for every }o\in U.
$
\end{theorem}

Thus, the fitted models asymptotically predict the distribution of learner observations correctly under every learner strategy. Appendix~\ref{app:opponent-recovery-proof} gives the proof.

\paragraph{Application to \SLME{}.}
Suppose the opponent is representable, meaning some mixture of library strategies is observationally equivalent to the opponent's strategy: there exists $\pi^*\in\Delta_M$ such that
\[
\sum_{m=1}^{M}\pi_m^*f_o(\sigma_{-i}^m)
=f_o(\sigma_{-i}^*)
\qquad\text{for every }o\in U.
\]
Then $\ell_n(\pi^*)=\ell_n(\sigma_{-i}^*)$.
Let $\widehat{\pi}_n:=\pi^{(k_n)}$ be the mixture after $k_n$ exact EM iterations from uniform weights. Since $\ell_n(\widehat{\pi})\geq\ell_n(\pi^*)$, Lemma~\ref{lem:slme-rate} gives
\[
\ell_n(\widehat\pi_n)
\geq \ell_n(\sigma_{-i}^*)
-\frac{n\log M}{k_n+1}.
\]
Thus, $k_n\to\infty$ ensures
Eq.~\eqref{eq:recovery-likelihood}, and the fitted mixtures
become asymptotically observationally equivalent to the opponent's strategy.
This conclusion requires the opponent to be
representable; it does not require the opponent's strategy
to be a library member.

\paragraph{Application to \OISR{}}
For any fixed pseudocounts $\lambda(I,a)\geq0$, a vanishing MAP optimization gap per play,
\[
\frac{J_n^*-J_n(\widehat{\sigma}_{-i,n})}{n}
\longrightarrow0
\qquad\text{almost surely},
\]
also implies Eq.~\eqref{eq:recovery-likelihood}.
To see this, write $y^*:=y_{-i}(\sigma_{-i}^*)$ and choose
a fixed realization-weight vector $y^{\mathrm{pos}}$
with positive entries.
For $n\geq2$, the vector 
$
y_n=(1-1/n)y^*+y^{\mathrm{pos}}/n\in Y_{-i}
$
has log likelihood at least $\ell_n(\sigma_{-i}^*)-2$.
Its action probabilities are at least $y^{\mathrm{pos}}_{Ia}/n$, so its log-prior contribution is bounded below by $-O(\log n)$.
Hence
\[
J_n^*\geq\ell_n(\sigma_{-i}^*)-O(\log n).
\]
Since $J_n(\sigma_{-i})\leq\ell_n(\sigma_{-i})$, an $o(n)$ MAP optimization gap implies the required likelihood bound.

Asymptotic observational equivalence thus needs only a vanishing MAP optimization gap per play, not coherence.
For \OISR{} with a fixed positive coherent prior and the initialization in Eq.~\eqref{eq:oisr-initialization}, Lemma~\ref{lem:oisr-rate} bounds the MAP optimization gap per play by $B_0/(n k_n)$ after $k_n$ exact EM iterations.
For a fixed finite game and prior, $\Gamma$ and $\tau$ are uniformly bounded, and therefore so is $B_0/n$ (Appendix~\ref{app:oisr-em-rate}). 
Choosing $k_n\to\infty$ ensures asymptotic observational equivalence.

\paragraph{Vanishing regret.}
Asymptotic observational equivalence also implies vanishing regret.

\begin{corollary}
\label{cor:recovery-and-regret}
Under the assumptions of Theorem~\ref{thm:opponent-recovery}, almost surely,
\[
R(\widehat{\sigma}_{-i,n})\longrightarrow0.
\]
\end{corollary}

\begin{proof}
The learner's terminal observation $o$ includes its utility, which we write $u_i(o)$.
Bounding utility prediction errors uniformly over learner strategies gives
\[
R(\widehat{\sigma}_{-i,n})
\leq
2\sum_{o\in U}|u_i(o)|
\left|f_o(\widehat{\sigma}_{-i,n})
-f_o(\sigma_{-i}^*)\right|
\longrightarrow0.\qedhere
\]
\end{proof}
\section{Empirical Evaluation}
\label{sec:experiments}
We compare \SLME{} and \OISR{} with three baselines in Leduc poker and two sizes of Deal or No Deal against three classes of opponents. In the offline setting (Section~\ref{subsection:static-fitting}), all methods see the same observed plays, numbering 0.01 to 100 times the opponent's number of information sets. In the online setting (Section~\ref{subsec:online}), the learner plays Leduc against the opponent, refits its model after each play, and best-responds to the refitted model in the next play.
Appendix~\ref{app:additional-experiments} reports the Kullback--Leibler (KL) divergence between the observation distributions of the opponent's strategy and each offline model, offline fitting time and peak memory, and \OISR{}'s sensitivity to its prior strength.
Our source code is available at \href{https://anonymous.4open.science/r/opponent-modeling-811E/}{https://anonymous.4open.science/r/opponent-modeling-811E/}.

\subsection{Experimental Setup}
\paragraph{Games, baselines, and strategy libraries} 
All games are implemented in OpenSpiel~\cite{lanctot2019openspiel}.
We denote the player who acts first by P1 and the second by P2.
Leduc poker has 468 information sets per player.
Players observe their own card, the public card, and all betting actions; the opponent's card is revealed only if neither player folds and the play ends in a showdown.
We use two sizes of Deal or No Deal~\cite{lewis2017deal}, each with up to three alternating proposals.
Each player's private valuation is drawn independently and uniformly from three types, each favoring a different item category.
Utilities are additive over allocated items, with zero payoff if no agreement is reached.
The small game contains one item per category and has 195 and 1,560 information sets for P1 and P2, respectively; the large game contains two items per category and has 2,190 and 59,130 information sets.
Players observe all proposals and responses, their own valuation, and their own payoff, but never the opponent's valuation.

We compare against Bayesian best response
(BBR)~\cite{southey2005bayes}, a finite-library baseline, and two library-free baselines: the EM estimation component of Mealing and Shapiro (MS-EM)~\cite{mealing2017em} and Ganzfried's \FMAP{}~\cite{ganzfried2026consistent}.
For \FMAP{}, we set every Dirichlet parameter to $2$ and, as in~\cite{ganzfried2026consistent}, use a realization-weight floor of $10^{-6}$ that keeps every realization weight positive.
Given the high computational costs of our \FMAP{} implementation, we evaluate it only in Leduc and small Deal or No Deal.\footnote{Our FMAP implementation uses Gurobi for projections. In large Deal or No Deal, some preliminary fits had not finished after six days.}

In each game, \SLME{} and \BBR{} share a predefined library of strategies that give every legal action positive probability. The Leduc library contains $48$ strategies. Each Deal or No Deal library contains $60$.

\paragraph{Opponent strategies.} 
We consider three classes, defined by their relationship to the strategy library.
A \emph{library member} uses one of the library's strategies. 
Each \emph{library mixture} has its own fixed subset of library strategies (up to 32 in Leduc and 24 in Deal or No Deal); before each play, it samples one uniformly and follows it throughout that play.
These two classes are \emph{representable} by the library. We verify that no library mixture is observationally equivalent to a single library strategy: the KL divergence between their observation distributions under a uniformly random learner (Eq.~\eqref{eq:observation-kl}, Appendix~\ref{app:library-distance}) is positive for every pair. 
We also use \emph{unrepresentable} opponents, whose strategies are not observationally equivalent to any mixture of library strategies. We generate each unrepresentable opponent's strategy by drawing its action probabilities at every information set from a Dirichlet distribution with all parameters equal to one and certify that it is unrepresentable (Appendix~\ref{app:library-distance}). Each opponent's strategy remains fixed throughout its interactions with the learner.

For each learner role, we evaluate the methods in 10 worlds, each specifying one opponent per class.
All methods face the same opponents within each world.

We use the \emph{relative dataset size} $c=n/|\mathcal I_{-i}|$, where $n$ is the number of observed plays and $|\mathcal I_{-i}|$ the opponent's number of information sets. Offline datasets target $c\in\{0.01,0.1,1,10,100\}$, rounding $n$ up to an integer; smaller datasets are prefixes of larger ones.

\paragraph{Evaluation metrics} 
We report regret: offline, the regret $R(\widehat{\sigma}_{-i,n})$ (Eq.~\eqref{eq:regret}) of a model fitted from $n$ observed plays, and online, the regret of the strategy the learner uses in each play.
We compute best responses using OpenSpiel~\cite{lanctot2019openspiel} and evaluate expected utilities against the opponent's strategy by full game-tree traversal.

To aid interpretation across payoff scales, we normalize by the regret of a uniform opponent model $\sigma_{-i}^{\mathrm{unif}}$, which assigns equal probability to every legal action at each information set. 
For each method, game, opponent class, dataset size, and learner role, we divide the average reported regret $R_w$ across the 10 worlds, indexed by $w$, by the corresponding average reference regret:
\begin{equation}
\bar R=\frac{\sum_w R_w}{\sum_w R_w(\sigma_{-i}^{\mathrm{unif}})}.
\label{eq:normalized-regret}
\end{equation}
A value of zero indicates optimal responses, one matches the reference, and values above one indicate worse performance.
We aggregate regrets before normalization to avoid unstable ratios in individual cases where a best response to the uniform opponent model already has near-zero regret.

Across all experiments, we compute pointwise 95\% bias-corrected and accelerated (BCa) bootstrap confidence intervals (CIs)~\cite{efron1987better} using 100,000 resamples of the 10 independent worlds within each learner role.

Appendix~\ref{app:kl} evaluates prediction accuracy using KL divergence, and Appendix~\ref{app:offline-resources} reports the fitting time and peak memory of offline fits, measured single-threaded on several CPU models and pooled across hosts.

The figures show each learner role separately. Unless noted otherwise, numbers in the text pool both roles, with the sums in Eq.~\eqref{eq:normalized-regret} running over the worlds of both roles.

\subsection{Offline Opponent Modeling}
\label{subsection:static-fitting}
\begin{figure}[!htbp]
\centering
\includegraphics{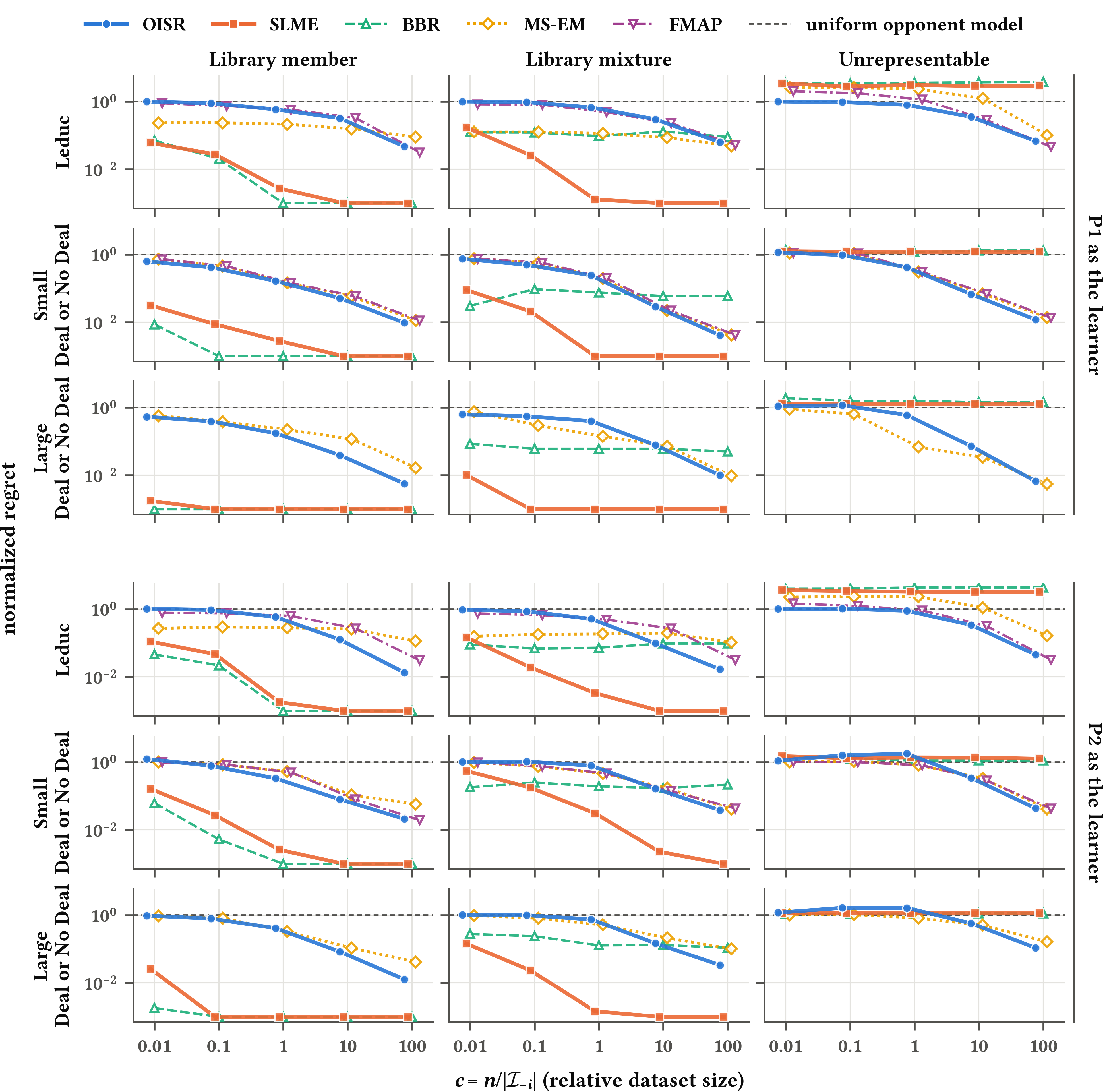}
\caption{Normalized regret of offline models (Eq.~\eqref{eq:normalized-regret}) for P1 (top) and P2 (bottom) as the learner. The pointwise 95\% bootstrap CIs have median width 0.087 normalized-regret units (max 2.51) for P1 and 0.138 (max 2.14) for P2.}
\label{fig:static-regret}
\end{figure}

We examine how the quality of responses to fitted opponent models changes with the number of observed plays $n$.
For each world, learner role, and opponent, all methods receive the first $n$ observations from the same sequence of plays, generated by a uniformly random learner.
Fig.~\ref{fig:static-regret} compares normalized regret (Eq.~\eqref{eq:normalized-regret}) across the three opponent classes and the dataset sizes.

\OISR{} uses the coherent prior of Appendix~\ref{app:oisr-prior} with strength $B_{\mathrm{prior}}=0.1$. 
For \SLME{}, \OISR{}, and \FMAP{}, we bound the remaining optimization gap using the maximum feasible improvement predicted by the objective's linearization~\cite{jaggi2013revisiting}.
Fitting stops when this bound per play falls below
$10^{-6}/\sqrt{n}$ or after $100{,}000$ optimization iterations; we evaluate the final iterate.\footnote{The optimization tolerance is met in all 900 \OISR{} fits, 734 of 900 \SLME{} fits, and 269 of 600 \FMAP{} fits. Another 53 \FMAP{} fits stop before reaching the tolerance because further projected gradient steps do not improve the objective.} 
\BBR{} computes posterior weights directly, while
MS-EM processes each play once.

\paragraph{Representable opponents.}
A suitable library enables effective responses from few observations, and \SLME{} retains this advantage when the opponent combines multiple library strategies (Fig.~\ref{fig:static-regret}, left two columns).
Across all three games, \SLME{} achieves normalized regret below $0.07$ at $c=0.1$ and below $0.01$ at $c=1$ against representable opponents.
Where evaluated, \OISR{} and \FMAP{} first achieve normalized regret below $0.07$ at $c=10$ or $100$, using roughly $100$--$1{,}000\times$ as many observations as \SLME{} at $c=0.1$; \MSEM{} remains above this threshold in Leduc at all tested dataset sizes.
Against library-member opponents, \BBR{} does even better, achieving normalized regret below $0.07$ already at $c=0.01$. Against library mixtures, however, its normalized regret stays between $0.07$ and $0.15$, higher than \SLME{}'s from $c=0.1$ on.
These results support using \SLME{} when observations are limited and domain knowledge supplies a small library of canonical strategies whose mixtures capture plausible opponent behavior.

\paragraph{Unrepresentable opponents.} 
Against unrepresentable opponents, \SLME{} and \BBR{} perform similarly to or worse than the uniform-model reference at all dataset sizes.
\OISR{} improves at larger dataset sizes, achieving normalized regret between $0.021$ and $0.058$ across the three games at $c=100$.
Using its compact representation, it achieves normalized regret comparable to \FMAP{}'s: $0.058$ versus $0.040$ in Leduc and $0.021$ versus $0.022$ in small Deal or No Deal.
\MSEM{} has lower regret than \OISR{} at some smaller dataset sizes in Deal or No Deal, but \OISR{} has lower regret in all three games at $c=100$.

\paragraph{Fitting time and peak memory.}
Figs.~\ref{fig:offline-time} and~\ref{fig:offline-memory} show fitting time and peak memory. On average across 300 matched fits per learner role in Leduc and small Deal or No Deal, \OISR{} and \SLME{} fit $269\times$ and $28\times$ faster than \FMAP{} for P1 as the learner and $32\times$ and $16\times$ faster for P2. They use $1.37\times$ and $1.36\times$ less peak memory than \FMAP{} for P1 and $1.11\times$ less each for P2. On average across 450 matched fits per learner role in all three games, \SLME{} fits $17\times$ slower than \BBR{} for P1 and $22\times$ slower for P2, and uses $1.51\times$ and $1.24\times$ more peak memory, respectively, showing the incurred computational cost of its repeated EM passes.

The computational costs of \FMAP{} and \MSEM{} scale poorly with game size. \FMAP{} optimizes over all of the opponent's sequences, whose number grows with the number of its information sets. For P1 as the learner, the opponent has 1,093 sequences in Leduc, 14,041 in small Deal or No Deal, and 1,655,641 in large Deal or No Deal; for P2, it has 1,093, 1,753, and 61,318. In large Deal or No Deal, this growth made \FMAP{} impractical: some fits had not finished after six days. \MSEM{} traverses the full game tree to initialize its counts and thus averages 2.76~GiB of peak memory per fit in large Deal or No Deal with just 592 observed plays for P1 as the learner, and 2.10~GiB with 22 plays for P2. On the same datasets, \SLME{} and \OISR{} average 119~MiB and 105~MiB of peak memory per fit for P1, and 103~MiB each for P2.

On average over the three games, the fitting time and peak memory of \SLME{} and \OISR{} increase with the number of observations for both learner roles. With each tenfold increase in $c$, the average fitting time grows by $2.8$--$16.9\times$ for \SLME{} and $3.1$--$8.0\times$ for \OISR{} for P1 as the learner, and by $5.2$--$11.2\times$ and $4.0$--$9.6\times$ for P2. The average peak memory grows by $1.43$--$2.32\times$ and $1.12$--$3.11\times$ for P1, and by $1.00$--$2.48\times$ and $1.00$--$2.41\times$ for P2. For \OISR{}, larger datasets typically expand the set of opponent sequences it must enumerate. On average across the three games, \OISR{} enumerates $0.8\%$, $19\%$, and $80\%$ of the full opponent sequence set required by \FMAP{} at $c=0.01$, $1$, and $100$ for P1 as the learner, and $1.0\%$, $22\%$, and $84\%$ for P2.

\subsection{Online Opponent Modeling: Leduc Poker}
\label{subsec:online}
\begin{figure}[!htbp]
\centering
\includegraphics{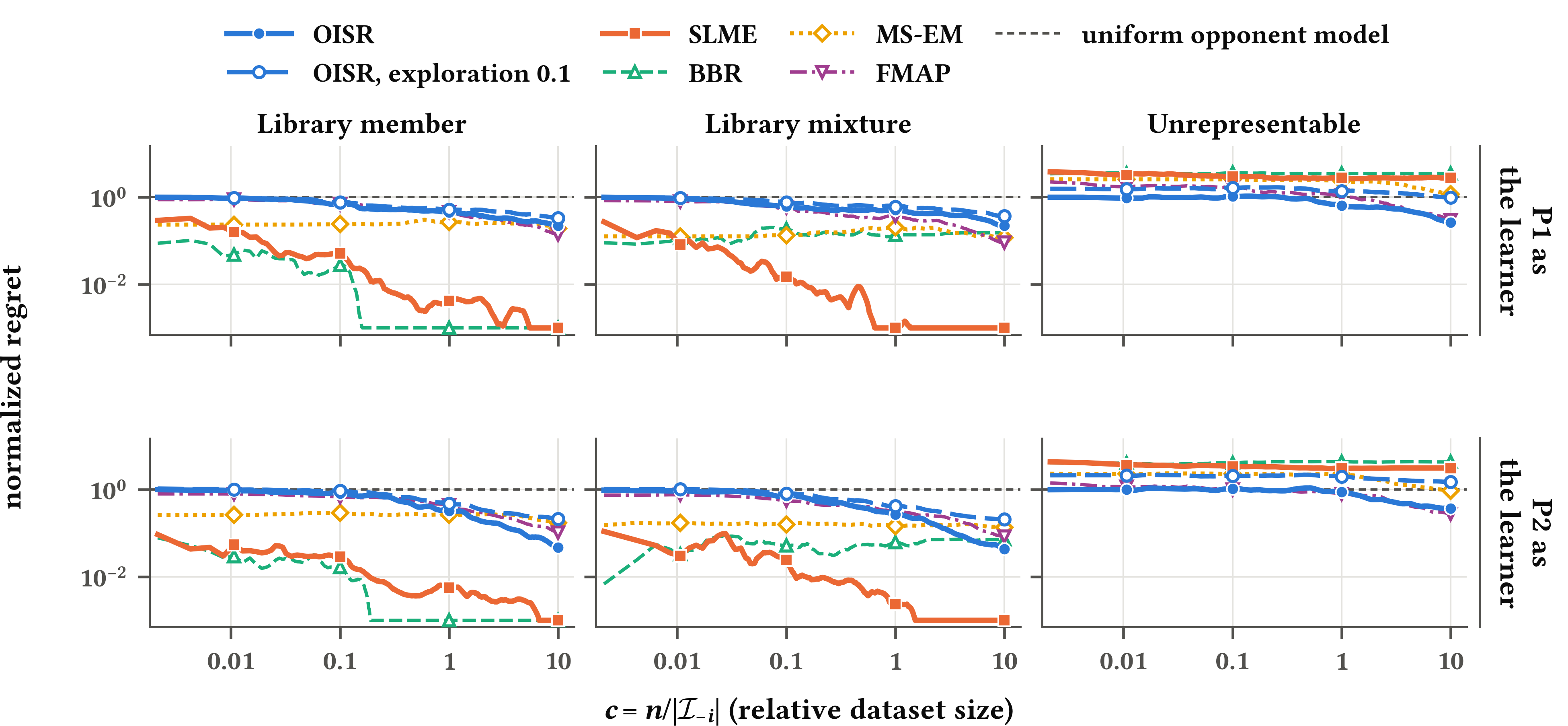}
\caption{Normalized regret per play during online play in Leduc (Eq.~\eqref{eq:normalized-regret}) for P1 (top) and P2 (bottom) as the learner. Curves are smoothed. The pointwise 95\% bootstrap CIs have median width 0.136 normalized-regret units (max 2.57) for P1 and 0.089 (max 2.13) for P2.}
\label{fig:online}
\end{figure}

We next evaluate opponent modeling during online play, where each method collects observations under its learner's current strategy; this setting follows prior work~\cite{ganzfried2026consistent,southey2005bayes}. 
The learner plays Leduc 4,680 times against a fixed opponent, starting with uniform play.
After each play, the learner updates its opponent model and computes a best response for the next play.
We also evaluate an \OISR{} exploration variant that mixes uniform play into the best response with probability $0.1$ at each learner decision.
Fig.~\ref{fig:online} reports the normalized regret of the learner's strategy used in each play for each learner role, pooled over 10 worlds.

For \SLME{}, \OISR{}, and \FMAP{}, each update uses the previous model for initialization and runs for at most 100 optimization iterations, stopping earlier when the per-play optimization-gap bound falls below $10^{-3}/\sqrt{n}$ after $n$ observed plays.
If an update fails, the model from the previous play is used.

The online results in Fig.~\ref{fig:online} reinforce the value of a suitable strategy library when observations are limited.
Against representable opponents,
\SLME{} achieves normalized regret below $0.01$ within 110 observed plays and averages at most $0.003$ over the 4,680 plays.

Against unrepresentable opponents, direct strategy
estimation is more effective than relying on the fixed library.
Using its compact representation, \OISR{} achieves
average normalized regret $0.49$, comparable to $0.54$ for \FMAP{} and substantially below $1.6$--$3.9$ for \MSEM{} and the library-based methods.
Uniform exploration with probability $0.1$ satisfies Eq.~\eqref{eq:statistical-coverage} but raises \OISR{}'s average normalized regret against unrepresentable opponents from $0.49$ to $1.4$, so over the tested horizon it provides no payoff benefit.

\section{Conclusion and Discussion}
\label{sec:conclusion}

We develop two complementary methods for opponent modeling from partial observations in imperfect-information games. \SLME{} fits mixtures of library strategies and is the method of choice when the opponent is representable by the library, whereas \OISR{} reconstructs a full strategy from a compact representation when no suitable library is known. Under sufficient exploration, \OISR{}'s model becomes asymptotically observationally equivalent to the opponent's strategy, and \SLME{}'s model does so whenever the opponent is representable.

Our analysis assumes a finite, two-player game and a fixed opponent strategy; with several unknown opponents, each play's observation probability becomes multilinear in the realization weights of their strategies and the objective may not be concave. Future work includes modeling adaptive opponents, exploration that balances information gathering and payoff, and automatic library construction.
\clearpage
\printbibliography[heading=bibintoc]
\clearpage
\appendix
\counterwithin{figure}{section}
\section{Supplemental Details for Section 3: Computational Cost}
\label{app:slme}
\label{app:slme-computation}

We count costs in calls to the game's rules, which never require the game tree, and in arithmetic operations. From a history, a call returns in unit time the acting player and its information set or, for a given action or \chance{} outcome, its legality, successor, and observation by the learner, plus a \chance{} outcome's probability; listing legal actions or outcomes, or computing a library strategy's distribution over them, takes time linear in their number. Arithmetic, $\log$, and $\exp$ cost one unit. Histories have at most $d_{\max}$ actions and \chance{} outcomes, and at most $b_{\max}$ actions or outcomes are legal at any history.

Summing likelihoods over hidden histories, for example over an opponent's possible unrevealed cards, is standard~\cite[Section~3.3]{southey2005bayes}; \SLME{} and \OISR{} do it without the game tree. Hashing each play's observation and the product of its learner action probabilities, which fix its known coefficients (Section~\ref{sec:problem-formulation}), groups the plays into types in expected time linear in the plays' total record length $N_R$. For each type, a depth-first search follows the recorded play. The search replays an event when the record identifies it exactly. Otherwise it tries each legal event and continues only when the resulting learner observations match a prefix of the record. It keeps terminal histories whose learner observations equal the full record. By perfect recall (Section~\ref{sec:problem-formulation}), every prefix of a compatible history passes this test, so the search finds exactly $Z_j$, and the \chance{} and learner probabilities along each give $c_{j,z}$. The search stores only its current path and the untried branches along it, so it never constructs the game tree. It takes $O(d_{\max}N_C)$ time, where $N_C$ counts its maximal paths over all types, ending at a kept history or at a mismatch.

\SLME{} accumulates each $q_j^m$, a sum over $Z_j$ (Section~\ref{sec:problem-formulation}), during the search in $O(Mb_{\max}d_{\max}N_H)$ time for the $N_H:=\sum_j|Z_j|$ histories found, each with at most $d_{\max}$ opponent actions. An EM iteration takes $O(|V|M)$ time (Section~\ref{sec:slme}), and under Lemma~\ref{lem:slme-rate}'s assumptions, $(\log M)/\varepsilon$ iterations from uniform weights suffice for a per-play optimization gap of at most $\varepsilon>0$. \SLME{} takes expected time $O(N_R+d_{\max}N_C+Mb_{\max}d_{\max}N_H+|V|M\log(M)/\varepsilon)$ and $O(N_R+b_{\max}d_{\max}+|V|M)$ space for the type keys, the search, and the table.
\section{Supplemental Details for Section 4: Optimization Guarantees and Computational Cost}
\label{app:oisr-details}

Throughout, $\sigma$ and $y_s$ abbreviate $\sigma_{-i}$ and $y_{-i,s}$. Except in Appendix~\ref{app:oisr-prior}, we fix the data ($n\geq1$) and assume the hypotheses of Theorem~\ref{thm:oisr-optimization}: a positive coherent prior and a positive entry in every row of $C$. A vector or strategy is \emph{positive} if all entries or action probabilities are.

\subsection{The Prior in Realization Weights}
\label{app:oisr-retained-prior}
We analyze EM in realization weights, where the log likelihood and, by coherence, the summed log prior are concave. Substituting $\sigma(a\mid I)=y_{Ia}/y_{\rho(I)}$ into the retained prior terms of Eq.~\eqref{eq:oisr-compact-objective} and collecting each $\log y_s$, with $\log y_\emptyset=0$, gives
\begin{equation}
J_n(y)-K=\ell_n(y)+\textstyle\sum_{s\in\widetilde{S}_{-i}}\kappa_s\log y_s,
\label{eq:oisr-retained-score}
\end{equation}
where $\kappa_\emptyset:=0$ and $\kappa_{Ia}:=\lambda(I,a)-\sum_{I'\in\widetilde{\mathcal{I}}_{-i}:\rho(I')=Ia}\lambda(I')$. Coherence (Eq.~\eqref{eq:oisr-coherence}) bounds $\lambda(I,a)$ below by the sum over all $I'$ with $\rho(I')=Ia$, retained or not, so every $\kappa_s\geq0$ and $J_n$ is concave in $y$.

The prior acts like imaginary, fully observed plays, allowed to stop at any retained sequence $s$, with likelihood proportional to $y_s$ and weight $\kappa_s$. A sequence \emph{extends} each of its prefixes, itself included; the imaginary plays whose sequence extends $Ia$ weigh
\begin{equation}
\textstyle\sum_{s\text{ extends }Ia}\kappa_s=\lambda(I,a),
\label{eq:oisr-prior-paths}
\end{equation}
by backward induction, as the plays continuing to a retained $I'$ with $\rho(I')=Ia$ weigh $\sum_{a'}\lambda(I',a')=\lambda(I')$, which $\kappa_{Ia}$ subtracts. Summing Eq.~\eqref{eq:oisr-prior-paths} over the opponent's first retained decisions shows that all imaginary plays weigh $\tau$; with the recorded plays, the total weight is $W:=n+\tau$.

\subsubsection{A positive coherent prior}
\label{app:oisr-prior}
We set the pseudocounts by depth, with bounds that certify coherence without listing the information sets. The depth $d_I$ of $I$ is the number of decisions the opponent makes before reaching $I$, and $D$ is the largest depth. Each information set of depth $d$ gets the same total pseudocount $\lambda_d>0$, split evenly: $\lambda(I,a):=\lambda_{d_I}/|A(I)|$, which is positive and needs only $d_I$ and $|A(I)|$, not the game tree. Let $A_d$ and $F_d$ bound, at depth $d$, the number of actions and the number of information sets $I'$ with $\rho(I')=Ia$ after each action $a$. For $d<D$, $\lambda(I,a)\geq\lambda_d/A_d$, and each of the at most $F_d$ information sets after $a$ has total pseudocount $\lambda_{d+1}$; at depth $D$, none follow. So $\lambda_d\geq A_dF_d\lambda_{d+1}$ for all $d<D$ makes the prior coherent. We set $\lambda_D:=B_{\mathrm{prior}}$, the prior strength, and, with the smallest bounds the rules allow, $\lambda_d:=1.1\,A_dF_d\lambda_{d+1}$ for $d<D$, or $\lambda_d:=B_{\mathrm{prior}}+1.1\,A_dF_d\lambda_{d+1}$ in Leduc.

\subsection{Initialization}
\label{app:oisr-initialization}
\label{app:oisr-coverage}
The endpoints carry the recorded plays, and the rate of Lemma~\ref{lem:oisr-rate} depends, through $\Gamma$, on how small their initial weights are, so the initialization of Eq.~\eqref{eq:oisr-initialization} mixes in feasible weights $x$ whose smallest endpoint weight, $1/\Gamma$, is as large as possible. Feasible weights, those in $\widetilde{Y}_{-i}$, are the nonnegative $y$ with $y_\emptyset=1$ that satisfy the flow constraints at the retained information sets: by the converse in Section~\ref{sec:problem-formulation}, every nonnegative vector with $y_\emptyset=1$ that satisfies the flow constraints is a realization-weight vector, and the prefixes of retained sequences are retained.

Each retained $I$ with $\rho(I)=s$ receives the full weight $y_s$ and splits it among its actions. So for every endpoint extending $s$ to get weight at least one, $s$ needs weight at least
\begin{equation}
r_s:=\max\Bigl\{\mathbf 1\{s\text{ is an endpoint}\},\ \textstyle\max_{I\in\widetilde{\mathcal{I}}_{-i}:\,\rho(I)=s}\sum_{a\in A(I)}r_{Ia}\Bigr\},
\label{eq:oisr-covering-requirements}
\end{equation}
where an empty maximum is zero. A backward pass over the retained sequences computes $r$ in linear time. An endpoint exists, as the rows of $C$ are nonzero, and Eq.~\eqref{eq:oisr-covering-requirements} makes $r$ at least one there and at its prefixes, so $r_\emptyset\geq1$. If a feasible $y$ has smallest endpoint weight $m$, backward induction with the flow constraints gives $y_s\geq m\,r_s$, so $m\leq1/r_\emptyset$. A forward pass attains this bound in linear time: from $x_\emptyset:=1$, at each retained $I$, it gives each action $a$ the weight $r_{Ia}/r_\emptyset$ and adds to one action the surplus $x_{\rho(I)}-\sum_ar_{Ia}/r_\emptyset$, which is nonnegative because, inductively, $x_{\rho(I)}\geq r_{\rho(I)}/r_\emptyset$. So $x$ is feasible with every endpoint weight at least $1/r_\emptyset$, and $\Gamma=r_\emptyset$. Induction on Eq.~\eqref{eq:oisr-covering-requirements} also bounds $r_s$ by the number of endpoints extending $s$, so $\Gamma\leq|S_{-i}|$ for every dataset, the uniform bound used in Section~\ref{sec:statistical-guarantees}.

As $\widetilde{\mathcal{I}}_{-i}\neq\emptyset$, the opponent's first decision on some compatible history is retained, so $\tau>0$ by positivity. So $y^{(0)}=(nx+\tau y^{\mathrm{def}})/W$, a mixture of feasible weights, is feasible and positive, with $y^{(0)}\geq(\tau/W)\,y^{\mathrm{def}}$ and, at endpoints, $y^{(0)}_s\geq nx_s/W\geq n/(W\Gamma)$.

\subsection{Proofs of Theorem~\ref{thm:oisr-optimization} and Lemma~\ref{lem:oisr-rate}}
\label{app:oisr-em-proof}
\subsubsection{Monotonicity and an optimal fixed point}
\label{app:oisr-em-monotonicity}
Jensen's inequality and the exact change in the retained log prior give the first inequality of Eq.~\eqref{eq:oisr-em-bound}, EM's ascent bound with the prior added~\cite[Section~2 and Theorem~1]{dempster1977em}. By the update in Eq.~\eqref{eq:oisr-em-update}, its right side equals $\sum_I(N^{(k)}(I)+\lambda(I))\,\KL\bigl(\sigma^{(k+1)}(\cdot\mid I)\,\Vert\,\sigma^{(k)}(\cdot\mid I)\bigr)$, which is positive unless the iteration leaves the model unchanged. Positive pseudocounts suffice, without coherence, and keep every iterate positive.

On the compact set of behavioral strategies, $J_n$ is upper semicontinuous and finite exactly at the positive strategies, so it attains $J_n^*$ at a positive strategy $\sigma^{\mathrm{opt}}$, which plays the prior modes at the non-retained information sets (Section~\ref{sec:oisr}) and has weights $y^{\mathrm{opt}}$ and expected counts $N^{\mathrm{opt}}$. By Eq.~\eqref{eq:oisr-em-bound}, an iteration from $\sigma^{\mathrm{opt}}$ cannot change it without exceeding $J_n^*$, so $\sigma^{\mathrm{opt}}(a\mid I)=(N^{\mathrm{opt}}(I,a)+\lambda(I,a))/(N^{\mathrm{opt}}(I)+\lambda(I))$.

\subsubsection{Global convergence}
\label{app:oisr-em-convergence}
To compare $\sigma^{(k)}$ with $\sigma^{\mathrm{opt}}$, let
\[
\mathcal D_k:=\textstyle\sum_{I\in\widetilde{\mathcal{I}}_{-i}}\bigl(N^{\mathrm{opt}}(I)+\lambda(I)\bigr)\,\KL\bigl(\sigma^{\mathrm{opt}}(\cdot\mid I)\,\Vert\,\sigma^{(k)}(\cdot\mid I)\bigr),
\]
which weights each retained $I$ by the expected weight of real and imaginary plays reaching it under $\sigma^{\mathrm{opt}}$ (Eq.~\eqref{eq:oisr-prior-paths}); it is nonnegative, and finite as $\sigma^{(k)}$ is positive. Let $\Delta_k:=J_n^*-J_n(\sigma^{(k)})$, and let $\beta(s):=\sum_jn_j\gamma_{j,s}+\kappa_s$ be the expected weight of real and imaginary plays whose complete opponent sequence is $s$, so that $\sum_s\beta(s)=W=n+\tau$; superscripts mark the model, and terms with $\beta^{\mathrm{opt}}(s)=0$ are zero. Then
\begin{equation}
\begin{aligned}
\mathcal D_{k+1}&\leq\textstyle\sum_s\beta^{\mathrm{opt}}(s)\log\bigl(\beta^{\mathrm{opt}}(s)/\beta^{(k)}(s)\bigr)\\
&\leq\textstyle\sum_jn_j\KL\bigl(\gamma^{\mathrm{opt}}_{j,\cdot}\,\Vert\,\gamma^{(k)}_{j,\cdot}\bigr)=\mathcal D_k-\Delta_k.
\end{aligned}
\label{eq:oisr-potential-decrease}
\end{equation}
For the equality, expanding each $\log(y^{\mathrm{opt}}_s/y^{(k)}_s)$ into action log-ratios and applying Eq.~\eqref{eq:oisr-prior-paths} and the fixed point gives $\sum_s\beta^{\mathrm{opt}}(s)\log(y^{\mathrm{opt}}_s/y^{(k)}_s)=\mathcal D_k$. By Eq.~\eqref{eq:oisr-retained-score}, $\Delta_k$ is this sum with each type's part $n_j\sum_s\gamma^{\mathrm{opt}}_{j,s}\log(y^{\mathrm{opt}}_s/y^{(k)}_s)$ replaced by $n_j\log(q_j(y^{\mathrm{opt}})/q_j(y^{(k)}))$; as $\gamma^{\mathrm{opt}}_{j,\cdot}$ sums to one and $y_s/q_j(y)=\gamma_{j,s}/C_{j,s}$ where $C_{j,s}>0$, the difference is the $n_j$-weighted divergence between each type's two posteriors over hidden sequences.

The second inequality follows by the log-sum inequality at each $s$, merging $n_j\gamma_{j,s}$ and the shared $\kappa_s\geq0$ (coherence). For the first, read $\beta/W$ as the distribution of a play's complete opponent sequence, generated one decision at a time: by Eq.~\eqref{eq:oisr-prior-paths} with the update in Eq.~\eqref{eq:oisr-em-update} and the fixed point, respectively, a play that reaches a retained $I$ continues with $a$ with probability $\sigma^{(k+1)}(a\mid I)$ under $\beta^{(k)}$ and $\sigma^{\mathrm{opt}}(a\mid I)$ under $\beta^{\mathrm{opt}}$, so the chain rule for relative entropy splits the first inequality's right side into $\mathcal D_{k+1}$ and nonnegative terms for stopping and for which retained information set comes next.

Summing Eq.~\eqref{eq:oisr-potential-decrease} over the first $k$ iterations gives $\sum_{l<k}\Delta_l\leq\mathcal D_0$; as the gaps never increase (Eq.~\eqref{eq:oisr-em-bound}), $\Delta_k\leq\mathcal D_0/k$ for $k\geq1$, which tends to zero and proves Theorem~\ref{thm:oisr-optimization}.\qed

\subsubsection{The rate}
\label{app:oisr-em-rate}
Since $\Delta_k\leq\mathcal D_0/k$, Lemma~\ref{lem:oisr-rate} follows once $\mathcal D_0\leq B_0$. As in the proof of Eq.~\eqref{eq:oisr-potential-decrease}, $\mathcal D_0=\sum_s\beta^{\mathrm{opt}}(s)\log(y^{\mathrm{opt}}_s/y^{(0)}_s)$. The real plays have total weight $n$ and lie on endpoints, where $y^{\mathrm{opt}}_s\leq1$ and $y^{(0)}_s\geq n/(W\Gamma)$ (Appendix~\ref{app:oisr-initialization}), so they contribute at most $n\log(W\Gamma/n)$. The part $\tau y^{\mathrm{def}}$ of Eq.~\eqref{eq:oisr-initialization} gives $y^{(0)}\geq(\tau/W)\,y^{\mathrm{def}}$, and the prior modes maximize $\sum_s\kappa_s\log y_s$, which by Eq.~\eqref{eq:oisr-prior-paths} equals the retained log prior ${\sum_{I,a}\lambda(I,a)\log\sigma(a\mid I)}$, so the imaginary plays, of total weight $\tau$, contribute at most $\tau\log(W/\tau)$. With $W=n+\tau$, the two bounds sum to $B_0$.\qed

Since $\log(1+u)\leq u$, we have $B_0/n\leq\log\Gamma+\tau/n+1$. For a fixed game and prior, the right side stays bounded as $n$ grows, as Section~\ref{sec:statistical-guarantees} requires, because $\Gamma\leq|S_{-i}|$ (Appendix~\ref{app:oisr-initialization}) and $\tau$ is at most the prior's total pseudocount at the opponent's first decisions.

\subsection{Computation}
\label{app:oisr-computation}
\label{app:oisr-pass-cost}
\paragraph{Escape actions.}
\label{app:oisr-escape}
To make each iteration cost $O(|V|+|E|+|\bar S|)$, \OISR{} represents only the sequences $\bar S$ along compatible terminal histories (Section~\ref{sec:oisr}), including the empty sequence, plus at most one escape weight per retained information set. At a retained $I$, the actions $a$ with $Ia\notin\bar S$, which no compatible history takes, enter $J_n$ only through their prior terms $\lambda(I,a)\log y_{Ia}$, and the flow constraint at $I$ only through their total weight: such an $Ia$ is no endpoint and precedes no retained information set, so $\kappa_{Ia}=\lambda(I,a)$ in Eq.~\eqref{eq:oisr-retained-score}. Where such actions exist, \OISR{} merges them into one escape action $e$ with their total pseudocount $\lambda(I,e)$. By Gibbs' inequality, splitting the escape weight in proportion to their pseudocounts maximizes their prior terms, making $J_n$ the merged objective plus a constant; so splitting the escape weights of a model fitted to the merged problem gives a model of the full MAP problem with the same optimization gap. The merged problem keeps all that Appendix~\ref{app:oisr-em-proof} uses of the prior (positivity, $\kappa\geq0$, Eq.~\eqref{eq:oisr-prior-paths}, and $\sum_s\kappa_s=\tau$), with $\kappa_{Ie}=\lambda(I,e)$. Merging preserves feasible endpoint weights and the totals $\lambda(I)$, hence $\Gamma$, $\tau$, and $B_0$, so Theorem~\ref{thm:oisr-optimization} and Lemma~\ref{lem:oisr-rate} hold for the merged problem, hence for the split models.

\paragraph{Costs.}
Preparation groups the plays and runs the search of Appendix~\ref{app:slme-computation}, in expected time $O(N_R+d_{\max}N_C)$. For each history $z\in Z_j$ found, \OISR{} adds $c_{j,z}$ to $C_{j,s_{-i}(z)}$ and enters the opponent's decisions along $z$ into hash tables of the retained information sets and of $\bar S$, in expected time $O(d_{\max})$, plus $O(b_{\max})$ to set the pseudocounts of each new information set. Each retained $I$ is followed by some $Ia\in\bar S$, and each history found has at most $d_{\max}$ opponent decisions, so $|\widetilde{\mathcal{I}}_{-i}|<|\bar S|\leq1+d_{\max}N_H$, and the merged problem has fewer than $2|\bar S|$ weights. Preparation, including the $O(|\bar S|)$ computation of $x$ and $y^{(0)}$, thus takes expected time $O(N_R+d_{\max}N_C+b_{\max}d_{\max}N_H)$. An EM iteration computes the $q_j$ and $\gamma_{j,s}$ over the nonzeros of $C$ in $O(|V|+|E|)$ time; a backward pass over $\bar S$ accumulates the expected number of recorded plays whose sequence extends each $s$, which at $s=Ia$ is $N(I,a)$, zero for escape actions; and a forward pass applies the update in Eq.~\eqref{eq:oisr-em-update} and recomputes the weights, in $O(|\bar S|)$ time. Beyond $O(|V|+|E|+|\bar S|)$ numbers for $C$, the counts, and the model, \OISR{} uses $O(N_R+b_{\max}d_{\max})$ space for type keys and the search.
\section{Supplemental Details for Section 5: Proof of \texorpdfstring{Theorem~\ref{thm:opponent-recovery}}{Theorem 2}}
\label{app:opponent-recovery-proof}
Let $r_n$ be the empirical distribution on $U$ of the observations in the first $n$ plays. For an opponent strategy $\sigma$, define the average observation distribution
\[
p_n^\sigma(o):=\left(\frac{1}{n}\sum_{t=1}^{n}y_{i,s_i(o)}(\sigma_i^t)\right)f_o(\sigma),
\]
with $s_i(o)$ the learner's sequence recorded in $o$. Write $p_n^*:=p_n^{\sigma_{-i}^*}$ and $\hat p_n:=p_n^{\widehat{\sigma}_{-i,n}}$. The strong law for bounded martingale differences gives $\|r_n-p_n^*\|_1\to0$ almost surely. Eq.~\eqref{eq:statistical-coverage} bounds $p_n^*(o)$ away from zero whenever $f_o(\sigma_{-i}^*)>0$, and no other observation can occur. Consequently, $\KL(r_n\,\|\,p_n^*)\to0$ almost surely.

The known learner factors cancel from likelihood ratios, giving
\[
\KL(r_n\,\|\,\hat p_n)=\KL(r_n\,\|\,p_n^*)+\frac{\ell_n(\sigma_{-i}^*)-\ell_n(\widehat{\sigma}_{-i,n})}{n}.
\]
Eq.~\eqref{eq:recovery-likelihood} therefore implies $\KL(r_n\,\|\,\hat p_n)\to0$. Pinsker's inequality and the triangle inequality yield $\|\hat p_n-p_n^*\|_1\to0$. Dividing each coordinate difference by the corresponding average learner realization weight, which is at least $\theta$ for all sufficiently large $n$ almost surely, proves the claim.\qed
\section{Supplemental Details for Section 6: Additional Experiments}
\label{app:additional-experiments}

\subsection{Certifying Unrepresentable Opponents}
\label{app:library-distance}

We compare opponent strategies by the KL divergence between the distributions they induce on the observation of a learner that plays uniformly at random,
\begin{equation}
\KL\p{q(\sigma_{-i})\,\|\,q(\sigma_{-i}')}=\sum_{j}q_j(\sigma_{-i})\log\frac{q_j(\sigma_{-i})}{q_j(\sigma_{-i}')},
\label{eq:observation-kl}
\end{equation}
where $j$ ranges over this learner's possible observations, each its own type (Section~\ref{sec:problem-formulation}). Observationally equivalent strategies induce the same distribution, so a positive divergence certifies that two strategies are not observationally equivalent.

Fix a learner strategy $\sigma_i$ that gives every legal action positive probability, such as the uniformly random learner of Eq.~\eqref{eq:observation-kl}, and let $j$ index its possible observations $o_j\in U$. Then $q_j(\sigma_{-i})=y_{i,s_i(o_j)}(\sigma_i)\,f_{o_j}(\sigma_{-i})$ with $y_{i,s_i(o_j)}(\sigma_i)>0$, and $q_j(\pi)=\sum_m\pi_mq_j^m$. For an opponent strategy $\sigma_{-i}^*$, let $p:=q(\sigma_{-i}^*)$ and
\[
\Phi(\pi):=\KL\p{p\,\|\,q(\pi)},\qquad d:=\min_{\pi\in\Delta_M}\Phi(\pi).
\]
Since \chance{}, $\sigma_i$, and every library strategy (Section~\ref{sec:experiments}) give every legal action positive probability, every $q_j^m$ is positive, and $\Phi$ is finite and continuous on $\Delta_M$; since $\Delta_M$ is compact, $\Phi$ attains its minimum $d$. The KL divergence vanishes only between equal distributions, and the learner factors are positive, so $\Phi(\pi)=0$ exactly when $\sum_m\pi_mf_o(\sigma_{-i}^m)=f_o(\sigma_{-i}^*)$ for every $o\in U$. Hence $d=0$ exactly when the opponent is representable (Section~\ref{sec:statistical-guarantees}), and a positive lower bound on $d$ certifies that it is not.

\begin{appendixlemma}[KL certificate]
\label{lem:kl-certificate}
For $\pi\in\Delta_M$, let $g_m(\pi):=\sum_jp_j\,q_j^m/q_j(\pi)$. Every $\pi\in\Delta_M$ satisfies
\[
d\geq L(\pi):=\Phi(\pi)+1-\max_mg_m(\pi),
\]
with equality when $\pi$ minimizes $\Phi$.
\end{appendixlemma}
\begin{proof}
$\Phi$ is convex because each $q_j(\pi)$ is linear in $\pi$. Its formula defines $\Phi$ on the open set where every $q_j(\pi)>0$, which contains $\Delta_M$. There $\partial\Phi/\partial\pi_m=-g_m(\pi)$, and for $\pi\in\Delta_M$, $\sum_m\pi_mg_m(\pi)=\sum_jp_j=1$. Since $\Phi$ lies above its tangent planes, every $\pi'\in\Delta_M$ satisfies
\[
\begin{aligned}
\Phi(\pi')&\geq\Phi(\pi)-\textstyle\sum_mg_m(\pi)(\pi'_m-\pi_m)\\
&=\Phi(\pi)+1-\textstyle\sum_m\pi'_mg_m(\pi)\geq L(\pi),
\end{aligned}
\]
because $\sum_m\pi'_mg_m(\pi)$ is a weighted average of the $g_m(\pi)$. Minimizing over $\pi'$ gives $d\geq L(\pi)$.

Now let $\pi$ minimize $\Phi$, and let $e_m$ be the vertex of $\Delta_M$ that selects member $m$. The segment $(1-t)\pi+te_m$ stays in $\Delta_M$ for $t\in[0,1]$. The derivative of $\Phi$ along it at $t=0$ is $-g_m(\pi)+\sum_r\pi_rg_r(\pi)=1-g_m(\pi)$, and at a minimizer it cannot be negative. Thus every $g_m(\pi)\leq1$. Their $\pi$-weighted average is $1$, so $\max_mg_m(\pi)=1$ and $L(\pi)=\Phi(\pi)=d$.
\end{proof}

With a uniformly random learner, running \SLME{}'s EM with $p_j$ in place of $n_j/n$, from uniform weights until $\max_mg_m(\pi)-1\leq10^{-3}\,\Phi(\pi)$, gives $L(\pi)>0$ for each of the 60 unrepresentable opponent strategies used in Section~\ref{sec:experiments}: at least $0.2358$ nats in Leduc, $0.6506$ in small Deal or No Deal, and $0.9477$ in large Deal or No Deal. This certifies that each is unrepresentable.

\subsection{KL Divergence from the Opponent's Strategy}
\label{app:kl}

We measure how well a fitted model $\widehat{\sigma}_{-i,n}$ predicts what the learner observes by the divergence of Eq.~\eqref{eq:observation-kl} from the opponent's strategy, $\KL(q(\sigma_{-i}^*)\,\|\,q(\widehat{\sigma}_{-i,n}))$, which is zero exactly when the model is observationally equivalent to the opponent's strategy. For \SLME{} and \BBR{}, $q_j(\widehat{\sigma}_{-i,n})$ is $q_j(\pi)$, with $\pi$ the fitted weights for \SLME{} and the posterior for \BBR{}.

\begin{figure}[p]
\centering
\includegraphics{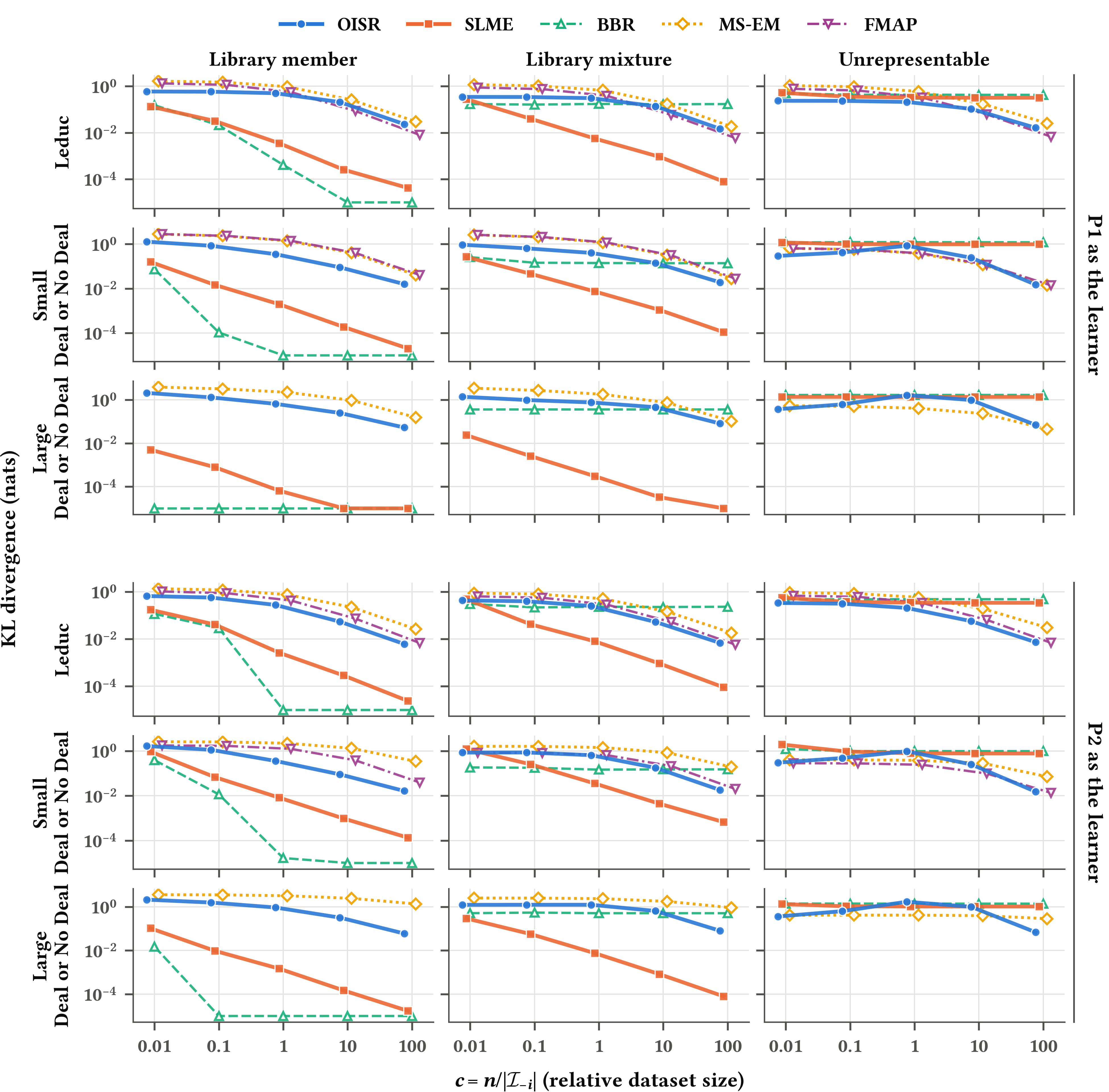}
\caption{KL divergence $\KL(q(\sigma_{-i}^*)\,\|\,q(\widehat{\sigma}_{-i,n}))$ (Eq.~\eqref{eq:observation-kl}) of each offline model from the opponent's strategy under a uniformly random learner, for P1 (top) and P2 (bottom) as the learner, averaged over 10 worlds. The pointwise 95\% bootstrap CIs have median width 0.037 nats (max 1.21) for P1 and 0.071 nats (max 1.44) for P2.}
\label{fig:kl}
\end{figure}

In Fig.~\ref{fig:kl}, as with regret (Fig.~\ref{fig:static-regret}), the library-based methods level off against unrepresentable opponents while \OISR{} keeps improving at large dataset sizes, and at $c=100$, \MSEM{}'s divergence is $1.7$--$13\times$ \OISR{}'s across all games and opponent classes. Unlike its regret, \SLME{}'s divergence keeps falling against library members and library mixtures after its responses are nearly optimal, by factors of 55--94 from $c=1$ to $c=100$. On average across games, opponent classes, and learner roles, \OISR{}'s KL divergence falls by factors of 1.07, 2.33, and 8.94 over the successive tenfold increases from $c=0.1$ to $c=100$.

\clearpage
\subsection{Offline Fitting Time and Peak Memory}
\label{app:offline-resources}

\begin{figure}[p]
\centering
\includegraphics{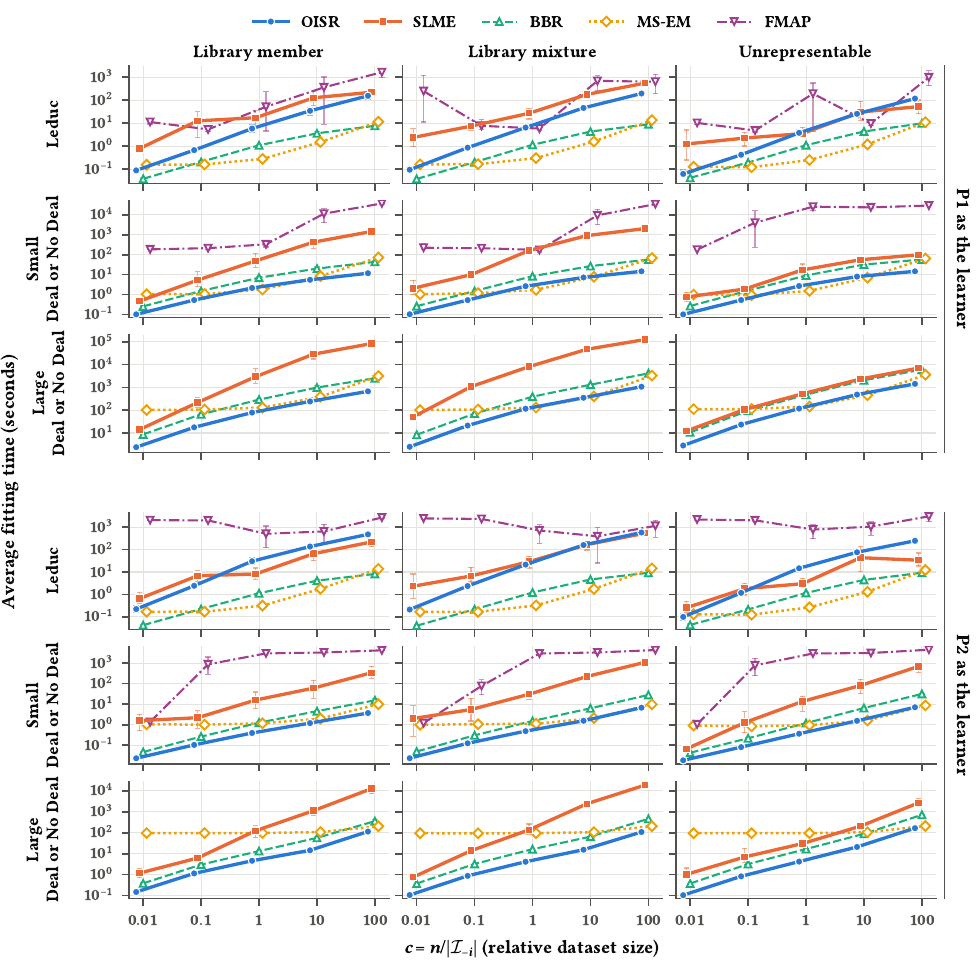}
\caption{Offline fitting time for P1 (top) and P2 (bottom) as the learner, averaged over 10 worlds. The pointwise 95\% bootstrap CIs, shown as bars, have median width 5.3~s (max 41,843~s) for P1 and 1.4~s (max 11,887~s) for P2.}
\label{fig:offline-time}
\end{figure}

\begin{figure}[p]
\centering
\includegraphics{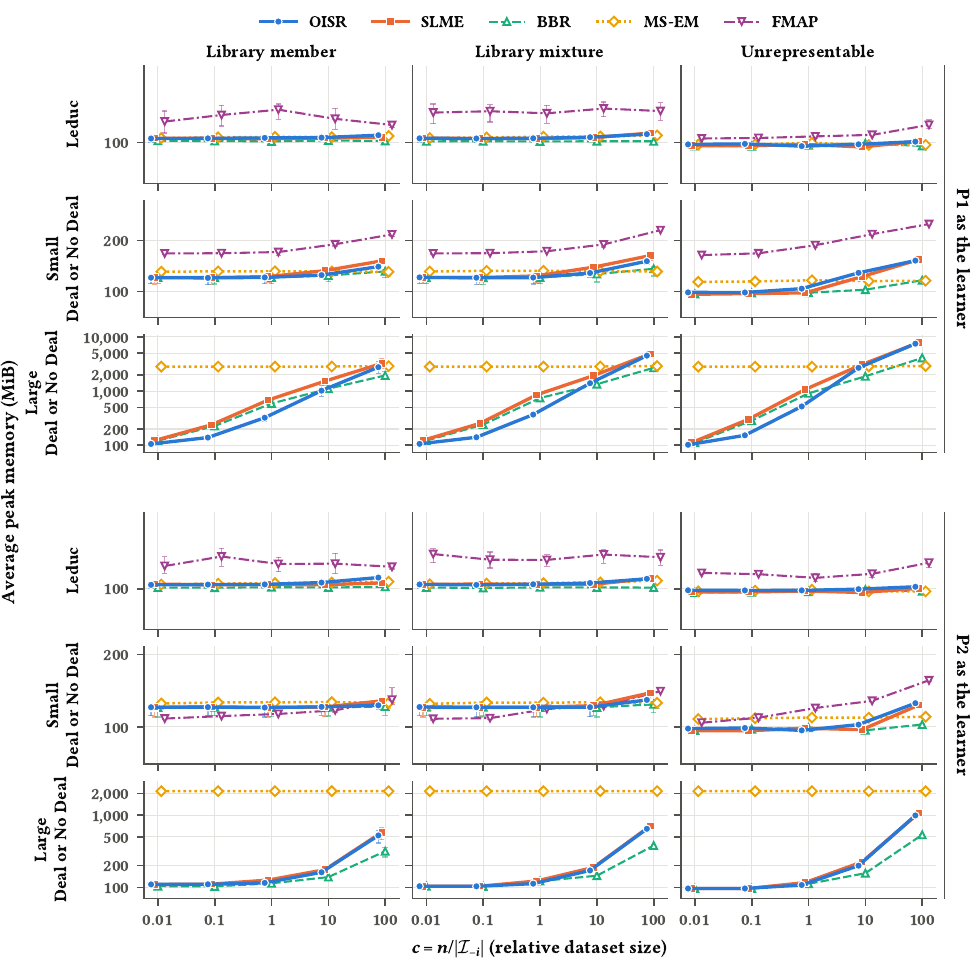}
\caption{Offline peak memory for P1 (top) and P2 (bottom) as the learner, averaged over 10 worlds. The pointwise 95\% bootstrap CIs, shown as bars, have median width 3.8~MiB (max 1,405~MiB) for P1 and 3.0~MiB (max 214~MiB) for P2.}
\label{fig:offline-memory}
\end{figure}

Figs.~\ref{fig:offline-time} and~\ref{fig:offline-memory} report single-threaded fits for the two learner roles. We report \FMAP{} only for Leduc and small Deal or No Deal.

In Fig.~\ref{fig:offline-time}, \OISR{}'s speed advantage over \FMAP{} grows with game size. On average over opponent classes and worlds at $c=100$, \OISR{} fits $7.0\times$ faster than \FMAP{} in Leduc for P1 as the learner and $5.3\times$ for P2, but $2{,}445\times$ and $739\times$ faster in small Deal or No Deal, where \FMAP{} averages 9.3 hours per fit for P1. In Deal or No Deal, \SLME{}'s fitting time grows faster with the data than \OISR{}'s: at $c=100$, depending on the learner role, \SLME{} takes $87$--$120\times$ as long as \OISR{} in small Deal or No Deal and $68$--$88\times$ in large Deal or No Deal, where its fits average 19.9 hours for P1 against 18 minutes for \OISR{}.

In Fig.~\ref{fig:offline-memory}, peak memory separates the methods mainly in large Deal or No Deal. In Leduc and small Deal or No Deal, on average over opponent classes and worlds, every method except \FMAP{} stays at or below 156~MiB at every dataset size; \FMAP{} reaches 232~MiB. In large Deal or No Deal, \MSEM{}'s peak memory barely changes with the data, 2.76 to 2.82~GiB for P1 as the learner and 2.10~GiB for P2, while that of \SLME{}, \OISR{}, and \BBR{} grows with it. At $c=100$, \OISR{} and \SLME{} average 4.80 and 5.20~GiB for P1, more than \MSEM{}, and 0.70 and 0.77~GiB for P2.

\subsection{Prior Strength}
\label{subsec:prior-ablation}

We vary $B_{\mathrm{prior}}$, which scales \OISR{}'s coherent prior: larger values more strongly favor uniform action probabilities (Appendix~\ref{app:oisr-prior}). We test $\{10^{-6},10^{-5},\ldots,10^3\}$ in Leduc using the offline dataset sizes and stopping criterion (Section~\ref{subsection:static-fitting}), pooling the three opponent classes over the 10 worlds; 5 of the 3,000 fits reach the iteration limit.

\begin{figure}[!tp]
\centering
\includegraphics{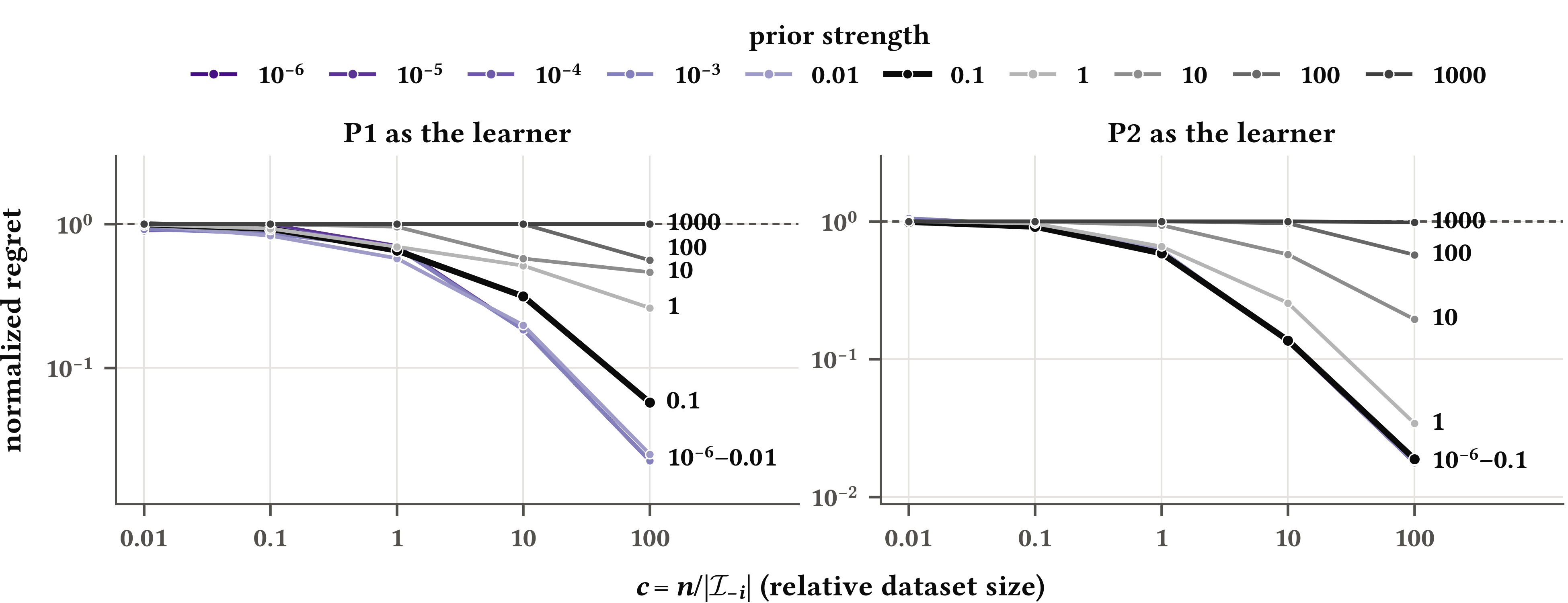}
\caption{\OISR{}'s normalized regret in Leduc by prior strength for P1 (left) and P2 (right) as the learner, pooled over the three opponent classes across the 10 worlds (Eq.~\eqref{eq:normalized-regret}). The pointwise 95\% bootstrap CIs have median width 0.053 normalized-regret units (max 0.247) for P1 and 0.055 (max 0.218) for P2.}
\label{fig:prior}
\end{figure}

Fig.~\ref{fig:prior} shows that stronger regularization limits the gains from additional observations. With five observed plays ($c=0.01$), all tested strengths give normalized regret near the uniform-model reference (0.96--1.00). The default $B_{\mathrm{prior}}=0.1$ attains pooled normalized regret within 5\% of the lowest among tested strengths for $c\leq1$, but at $c=100$ its normalized regret is 0.039, compared with approximately 0.020 for $B_{\mathrm{prior}}\leq10^{-3}$. A smaller fixed strength $B_{\mathrm{prior}}=0.01$ attains pooled normalized regret within 8\% of the lowest at every tested dataset size.

\clearpage

\clearpage
\end{document}